\documentclass[12pt]{article}

\usepackage[dvips]{graphicx}
\usepackage{epsfig}
\usepackage{amsmath,amsfonts,amssymb,amsthm}
\usepackage{verbatim}
\usepackage{psfrag}
\usepackage{bm}
\usepackage{bbm}
\usepackage[square,comma,sort&compress,numbers]{natbib}
\usepackage{color}
\usepackage{slashed}

\usepackage{epsf,epsfig}
\usepackage{graphics}

\def\be{\begin{equation}}
\def\ee{\end{equation}}
\def\bea{\begin{eqnarray}}
\def\eea{\end{eqnarray}}

\begin{document}
\thispagestyle{empty}

\begin{flushright}
{
\small
TTK-10-44\\
ZU-TH~09/10\\
%10mm.nnnn [xxx]\\[0.2cm]
July 27, 2010
}
\end{flushright}

\vspace{0.4cm}
\begin{center}
\Large\bf\boldmath
Flavoured Leptogenesis in the CTP Formalism
\unboldmath
\end{center}

\vspace{0.4cm}

\begin{center}
{Martin Beneke,
Bj\"orn~Garbrecht,
Christian Fidler,
Matti Herranen}\\
\vskip0.3cm
{\it Institut f\"ur Theoretische Teilchenphysik und Kosmologie,\\ 
RWTH Aachen University,\\
D--52056 Aachen, Germany}\\
\vskip.3cm
and\\
\vskip.3cm
{Pedro Schwaller}\\
\vskip0.3cm
{\it Institut f\"ur Theoretische Physik,\\
 Universit\"at Z\"urich, CH--8057 Z\"urich, Switzerland}
\end{center}

\vspace{1.3cm}
\begin{abstract}
\vspace{0.2cm}\noindent
Within the Closed Time Path (CTP) framework, we derive
kinetic equations for particle distribution functions
that describe leptogenesis in the presence of several lepton flavours.
These flavours have different
Standard-Model Yukawa couplings, which induce flavour-sensitive scattering 
processes and thermal dispersion relations.
Kinetic equilibrium, which
is rapidly established and maintained via gauge interactions, allows to
simplify these equations to kinetic equations for the matrix of
lepton charge densities.
In performing this simplification, we notice that the rapid
flavour-blind gauge interactions damp the flavour oscillations of the
leptons. Leptogenesis turns out to be in the parametric regime
where the flavour oscillations are overdamped and flavour decoherence
is mainly induced by flavour sensitive scatterings. We solve the
kinetic equations for the lepton number densities numerically and
show that they interpolate between the unflavoured and the fully flavoured
regimes within the intermediate parametric region, where neither of these 
limits is applicable.

\end{abstract}

\newpage
\setcounter{page}{1}

%\newpage
\allowdisplaybreaks[2]

\section{Introduction}

Calculations of the baryon asymmetry of the Universe require the 
description of $CP$-violating processes in a finite-density background. 
Conventionally, semi-classical Boltzmann equations are employed, where one 
computes the evolution of classical particle number densities in the 
presence of scattering processes, that are given by matrix elements from 
the in-out formalism. Since $CP$ violation in these
matrix elements is a higher order effect that is typically induced by loop 
diagrams, where particles in the loop are kinematically allowed to be 
on shell, this approach bears notorious difficulties concerning the 
correct counting: Is a certain process higher order or is it a combination 
of two leading order processes, that are already accounted for in the 
Boltzmann equations? For the $CP$-violating processes in 
baryogenesis, a way of addressing this caveat is the method of real 
intermediate state (RIS) subtraction~\cite{Kolb:1979qa}.

A more direct and systematic way to deal with this counting problem is
to avoid the detour via
the in-out matrix elements and instead to formulate the problem directly in
terms of Green functions. A method that achieves this is given in terms
of the Closed Time Path (CTP) or in-in 
formalism~\cite{Schwinger:1960qe,Keldysh:1964ud},
which has been successfully applied to
various models of baryogenesis including also a realistic scenario
of leptogenesis~\cite{Buchmuller:2000nd,De Simone:2007rw,Garny:2009rv,Garny:2009qn,Anisimov:2010aq,Beneke:2010wd,Garny:2010nz}
. The set of Green functions that are computed in the CTP framework encompasses
both, the spectral and the statistical information of the system. To leading
accuracy in a weakly interacting situation, it is often sufficient to approximate
the spectral functions by on-shell $\delta$-functions, while the statistical
contributions encode the distribution functions of quasi-particles. In this way,
the kinetic equations that govern leptogenesis have been recovered systematically
and additional corrections due to the full quantum statistical distributions
of right-handed neutrinos and the leptons and Higgs boson of the Standard Model
have been derived~\cite{Beneke:2010wd}.

The discussion in Ref.~\cite{Beneke:2010wd} is for a single flavour
of Standard Model leptons $\ell$, which is appropriate in those situations
where the different Standard Model Yukawa couplings  of these flavours are negligible,
because an asymmetry is only produced
for one particular linear combination of the lepton flavours. However,
at smaller temperatures, when the lepton Yukawa couplings $h$ of the Standard Model
approach equilibrium, the flavour degeneracy is broken and effects of flavour
have to be taken into account~\cite{Endoh:2003mz,Pilaftsis:2005rv,Abada:2006fw,Nardi:2006fx}. The derivation of a set of kinetic equations from 
non-equilibrium quantum field theory, which covers both the unflavoured 
and fully flavoured regimes, and is valid in between, is still missing 
up to now, and is the subject of this paper.

A simple model that encompasses the salient features of flavoured 
leptogenesis and which we consider in the present work is specified by the
Lagrangian
\begin{align}
\label{TheLagrangian}
{\cal L}=&
\frac12 \bar\psi_{Ni}({\rm i}\partial\!\!\!/-M_i)\psi_{Ni}
+\bar\psi_{\ell a}
{\rm i}\partial\!\!\!/
\psi_{\ell a}
+\bar\psi_{{\rm R} b}
{\rm i}\partial\!\!\!/
\psi_{{\rm R} b}
+(\partial^\mu \phi^\dagger)(\partial_\mu \phi)
\\\notag
&-Y_{ia}^*\bar{\psi}_{\ell a}\tilde \phi \psi_{Ni}
-Y_{ia}\bar{\psi}_{Ni}\tilde\phi^\dagger \psi_{\ell a}
-h_{ab}\phi^\dagger \bar\psi_{{\rm R}a} P_{\rm L} \psi_{\ell b}
-h_{ab}^*\phi \bar\psi_{\ell b} P_{\rm R} \psi_{{\rm R} a}
\,,
\end{align}
where $\psi_{Ni}$ is the four-component Majorana spinor representing
the right handed singlet neutrino $N_i$, $\psi_{\ell a}$ is the spinor
for the ${\rm SU}(2)_{\rm L}$ doublet of left handed Standard Model 
leptons $\ell_a$, where $a$ is the flavour index, and $\psi_{{\rm R}a}$ are 
the corresponding charged right handed leptons. The ${\rm SU}(2)_{\rm L}$ 
doublet of Higgs fields is represented by $\phi$, and we define 
$\tilde \phi = (\epsilon \phi)^\dagger$ with $\epsilon$ the 
antisymmetric $2\time 2$ matrix in the ${\rm SU}(2)_{\rm L}$ indices 
with $\epsilon_{12}=1$. For simplicity, we assume $i=1,2$. 
The generalisation to more than two
right handed neutrinos $N_i$ is straightforward.

Single-flavour calculations for leptogenesis can be easily generalised
to the multi-flavour situation, provided there is a flavour basis in which one
can express the lepton-number densities of the several flavours and at the 
same time neglect possible correlations between the different flavours.
When the lepton Yukawa couplings are fully in equilibrium, the appropriate 
flavour basis is where these
couplings $h$ are diagonal. In contrast, in the unflavoured regime
the appropriate basis is determined by the linear combination, in which the
lepton asymmetry is produced.

These considerations raise the following questions: First, is there a kinetic
equation that is manifestly covariant under the choice of the flavour basis? 
And second, can such an equation also deal with the intermediate regime 
where the couplings $h$ are not yet in full equilibrium but are 
non-negligible at the same time? The latter point is of particular 
importance since leptogenesis is a process that
takes a finite amount of time while temperature is decreasing,
such that the flavour-sensitive interactions
may be out of equilibrium initially, but equilibrate at later times. Clearly,
in order to address this question, one has to promote the set of lepton number
distributions within a certain flavour basis to a matrix, that also allows
for off-diagonal flavour correlations. In a heuristic approach, by appealing 
to the Hamiltonian evolution of a density matrix, such a set of equations 
has been proposed~\cite{Abada:2006fw}, and numerical solutions to these 
equations have been obtained~\cite{De Simone:2006dd}.

On the other hand, it is clear that the Green functions within the CTP 
formalism allow for the possibility of off-diagonal correlations in a 
straightforward way. Indeed, once the appropriate Kadanoff-Baym equations for
unflavoured leptogenesis are known~\cite{Beneke:2010wd}, the multi-flavour
generalisation is easily written down. However, the Kadanoff-Baym equations
are a coupled set of integro-differential equations that needs to be subjected 
to some analytic approximations before solving it numerically. 
Key simplifications are first
the gradient expansion, which effectively is an expansion in terms of time
derivatives, in deviations of the distribution functions from equilibrium, 
and in the small coupling constants. Second, one makes use of a separation 
of time scales between the interactions induced by the 
small Yukawa couplings $Y$ and $h$ and  the faster gauge interactions, that
maintain kinetic equilibrium and induce lepton-antilepton pair creation 
and annihilation processes. This allows to describe the distribution 
functions by generalised chemical potentials.

The application of these strategies of simplification to the multi-flavour 
case constitutes the main body of the present work. In 
Section~\ref{section:flavlep}, we set up the multi-flavour Kadanoff-Baym 
equations and show that the flavour-dependent dispersion relations for the 
leptons, that are induced by the right handed neutrino and Yukawa couplings 
$Y$ and $h$, respectively, give rise
to commutator terms in the kinetic equations which are also characteristic
of flavour oscillations in the presence of tree-level mass terms. 
In Section~\ref{sec:kinetic}, we make use of the short time scale 
for kinetic equilibration in order to express the lepton densities in 
terms of a generalised chemical potential. We show that the gauge 
interactions in addition to enforcing kinetic equilibrium damp
the flavour oscillations and that parametrically, flavoured leptogenesis is 
in the overdamped regime. The processes that turn out to dominate the decay
of the off-diagonal flavour correlations are the flavour-sensitive
three-body scatterings through
the couplings $h$.

Let $q_{\ell ab}$ denote the Hermitian matrix of lepton charge densities, 
including flavour off-diagonal correlations, which is defined more 
precisely below. The main result of the present work is the
kinetic evolution equation
\begin{align}
\label{kin:eq}
\frac{\partial q_{\ell ab}}{\partial \eta}
=
\sum\limits_{c}\left[
q_{\ell ac}\Xi_{cb}
-\Xi_{ac}q_{\ell cb}
-W_{ac} q_{\ell cb}
-q_{\ell ac}W_{cb}
\right]
+ 2 S_{ab}
-\Gamma_{\ell ab}^{\rm fl}
\,.
\end{align}
In this equation, $\eta$ denotes the conformal time,
which is related to the physical comoving time through $d t= a(\eta) d \eta$,
where $a(\eta)$ is the scale factor of the expanding Universe.
The matrices $W$, $S$ and $\Gamma_\ell^{\rm fl}$ are all Hermitian.
 Lepton number
and $CP$-violating source terms are represented by $S$, $W$ encompasses
the washout rates for the various lepton flavours, and
$\Gamma_\ell^{\rm fl}$ is the matrix of
flavour-sensitive damping rates. In case we impose that
$q_{\ell ab}$ is evaluated in the basis of mass eigenstates of leptonic
quasi-particles, the anti-Hermitian matrix $\Xi$ compensates for
possible time-dependent flavour rotations.
Numerical solutions to this equation are presented in 
Section~\ref{section:numerics}, and
it is shown that it indeed interpolates between the fully flavoured and
the unflavoured regimes. 

We conclude in Section~\ref{sec:conclude}. 
Appendix~\ref{appendix:finitewidth} contains a discussion of the 
constraint and mass-shell equations including finite width effects, 
and the contribution of the right-handed neutrino to the 
thermal mass of the leptons is calculated in Appendix~\ref{appendix:thmass}.
\section{Flavoured Leptons}
\label{section:flavlep}

The usual strategy for deriving kinetic equations is to
decompose the Schwinger-Dyson equations on the
CTP into equations for the retarded and advanced propagators and the Kadanoff-Baym
equations. In the weak coupling limit, one can solve the latter
when approximating the particle densities as purely on-shell. This
corresponds to taking the particle width to zero.

In this Section, we reiterate these arguments for the left-handed leptons.
As an extension of earlier approaches~\cite{Konstandin:2004gy,Konstandin:2005cd,Cirigliano:2009yt},
where mass terms are introduced at tree-level,
we describe here the dynamics that arises from thermal one-loop corrections to the dispersion
relations, which are mediated by flavour-blind gauge interactions as well as flavour-sensitive
Yukawa interactions.

\subsection{Schwinger-Dyson Equations}

We employ here the notations and conventions for the CTP formalism and
the gradient expansion that are explained in more detail
within Ref.~\cite{Beneke:2010wd,Prokopec:2003pj,Prokopec:2004ic}.
The Schwinger-Dyson equation for the flavoured left-handed lepton propagator is
\begin{align}
{\rm i}\partial\!\!\!/_u S_{\ell ab}^{fg}(u,v)
=f \delta^{fg}\delta_{ab} \delta^4(u-v) P_{\rm R}
+\sum\limits_h\int d^4 w {\Sigma\!\!\!\!/}^{fh}_{\ell ac}(u,w) S_{\ell cb}^{hg}(w,v)
\,,
\end{align}
which can be more compactly written as~\cite{Prokopec:2003pj,Prokopec:2004ic}
\begin{align}
{\rm i}\partial\!\!\!/ S_\ell^{fg}=f\delta^{fg} \delta P_{\rm R} +\sum\limits_h{\Sigma\!\!\!\!/}^{fh}_{\ell} \odot S^{hg}_{\ell}
\,,
\end{align}
where the symbol $\odot$ denotes a convolution and $\delta$ the delta 
function, $f,g,h=\pm$ are the CTP indices
and $a,b,c$ flavour indices.
The CTP structure can be decomposed into equations for the retarded and
advanced propagators and a
Kadanoff-Baym equation,
\begin{subequations}
\label{CTP:equations}
\begin{align}
{\rm i}\partial\!\!\!/S_\ell^{A,R}&=
\delta P_{\rm R}+\Sigma_{\ell}^H\odot S_\ell^{A,R}
\pm {\rm i}{\Sigma\!\!\!\!/}^{\cal A}_\ell \odot S_\ell^{A,R}\,,
\\
{\rm i}\partial\!\!\!/S_\ell^{<,>}&=
{\Sigma\!\!\!\!/}_{\ell}^H\odot S_\ell^{<,>}
+ {\Sigma\!\!\!\!/}^{<,>}_\ell \odot S_\ell^{H}
+\frac 12\left(
\Sigma_\ell^> \odot S_\ell^<- \Sigma^<_\ell \odot S^>_\ell
\right)\,.
\end{align}
\end{subequations}
Here and in what follows, we make use of the definitions
\begin{align}
G^<=G^{+-}\,,\quad
G^>=G^{-+}\,,\quad
G^T=G^{++}\,,\quad
G^{\bar T}=G^{--}
\end{align}
and the combinations
\begin{align}
\label{CTP:combinations}
G^A=G^T-G^>=G^<-G^{\bar T}\quad&\textnormal{(advanced)}\,,
\\\nonumber
G^R=G^T-G^<=G^>-G^{\bar T}\quad&\textnormal{(retarded)}\,,
\\\nonumber
G^H=\frac12(G^R+G^A)\quad&\textnormal{(Hermitian)}\,,
\\\nonumber
G^{\cal A}=\frac1{2\rm i}(G^A-G^R)=\frac{\rm i}2(G^>-G^<)\quad&\textnormal{(anti-Hermitian, spectral)}\,,
\end{align}
where $G$ may stand for any two-point function on the CTP,
in particular the propagators and self-energies.

The Wigner transformation is defined as a Fourier transformation of a 
two-point function with respect to the relative coordinate, while keeping 
the average coordinate fixed:
\begin{align}
G(k,x)=\int d^4 r\, {\rm e}^{{\rm i}k \cdot r}G(u,v)\,,
\quad
\textnormal{where}\;\;
r=u-v\;\; \textnormal{and}\;\;
x=\frac{u+v}2\,.
\end{align}
It provides a separation between the typical energy scale (which is the
temperature $T$ in the present case) and the small macroscopic inverse 
time-scale, that governs the evolution of state parameters (particle number 
and charge distributions in the present case). The ratio of these two time 
scales is then used to define the gradient expansion.
In Wigner space, Eqs.~(\ref{CTP:equations}) take the form
\begin{subequations}
\label{CTP:eq:Wigner}
\begin{align}
{\rm e}^{-{\rm i}\diamond}
\left\{
k\!\!\!/-{\Sigma\!\!\!/}_\ell^H\mp{\Sigma\!\!\!/}_\ell^{\cal A}
\right\}
\left\{
S_\ell^{A,R}
\right\}&=P_{\rm R}\,,
\\
\label{KB:Wigner}
{\rm e}^{-{\rm i}\diamond}
\left\{
k\!\!\!/-{\Sigma\!\!\!/}^H_\ell
\right\}
\left\{
S_\ell^{<,>}
\right\}
-{\rm e}^{-{\rm i}\diamond}
\left\{
{\Sigma\!\!\!/}^{<,>}_\ell
\right\}
\left\{
S_\ell^H
\right\}
&=\frac12{\rm e}^{-{\rm i}\diamond}
\left(
\left\{
{\Sigma\!\!\!/}^{>}_\ell
\right\}
\left\{
S_\ell^<
\right\}
-
\left\{
{\Sigma\!\!\!/}^{<}_\ell
\right\}
\left\{
S_\ell^>
\right\}
\right)\,,
\end{align}
\end{subequations}
where
\begin{align}
\diamond\,\{A(k,x)\}\{B(k,x)\}
=\frac12\left(
[\partial_{(x)}^\mu A(k,x)]\, \partial_{(k)\mu}B(k,x)
-
[ \partial_{(k)\mu}A(k,x)]\,\partial_{(x)}^\mu B(k,x)
\right)
\,.
\end{align}
Eqs.~(\ref{CTP:eq:Wigner}) correspond to an infinite tower of 
integro-differential equations. Approximate solutions can be obtained 
within the scheme of gradient expansion~\cite{Calzetta:1986cq}.
In the context of leptogenesis, gradients occur due to the
deviation of particle number distributions
from equilibrium, which is induced by the Hubble expansion of the Universe.
At the same time,  for a sizeable lepton asymmetry
to occur, it is crucial, that during leptogenesis, the rate $|Y_{1a}|^2 T$
is not very different from expansion rate $H$. Similarly, within this
paper we are particularly interested in the parametric region where the 
$\tau$-lepton Yukawa coupling $h_\tau$ relates to $H$ as 
$h_\tau^2 T \sim H$, since
otherwise we are either in the fully flavoured or in the unflavoured regime,
which can be described by conventional approaches. In the present context, gradient
expansion is therefore understood not only as
an expansion in time-derivatives,
but also as a perturbative expansion in $Y$ and $h$.

Performing the expansion of Eqs.~(\ref{CTP:eq:Wigner})
up to first order in gradients, we obtain
\begin{subequations}
\begin{align}
\label{polemass:gradexp}
\left(k\!\!\!/-{\Sigma\!\!\!\!/}_\ell^H\mp{\Sigma\!\!\!\!/}_\ell^{\cal A}
\right)
S_\ell^{A,R}&=P_{\rm R}\,,\\
\label{KB:gradexp}
\frac{\rm i}{2}\partial\!\!\!/S_\ell^{<,>}
+(k\!\!\!/-{\Sigma\!\!\!/}_\ell^H)S_\ell^{<,>}
-{\Sigma\!\!\!\!/}^{<,>}_\ell S_\ell^H
&=
\frac12
\left(
{\Sigma\!\!\!/}^{>}_\ell
S_\ell^<
-
{\Sigma\!\!\!/}^{<}_\ell
S_\ell^>
\right)
\,.
\end{align}
\end{subequations}
The derivative term $\frac{\rm i}2\partial\!\!\!/ S_\ell^{A,R}$ would  
contribute to Eq.~(\ref{polemass:gradexp}) only at second order in gradients, 
since the retarded and
advanced propagators do not depend on the particle distribution functions at
tree level. Damping occurs explicitly through the collision term
on the right hand side of Eq.~(\ref{KB:gradexp}) 
(for a detailed discussion, see Ref.~\cite{Garbrecht:2008cb}).
At the same time, damping is contained in the
${\Sigma\!\!\!\!/}_\ell^{\cal A}$ term in Eq.~(\ref{polemass:gradexp}), 
as it is well known from linear response theory.

In order to take account of the dilution of particles due to the expansion of
the Universe, we follow Ref.~\cite{Beneke:2010wd}.
We first observe that after appropriate field redefinitions
and the rescaling $M_i\to a(\eta) M_i$, the Lagrangian~(\ref{TheLagrangian}) describes
the fields in the background of a flat Friedmann-Robertson-Walker Universe
in conformal coordinates defined by the metric
\begin{align}
\notag
g_{\mu\nu}=a^2(\eta)\,{\rm diag}(1,-1,-1,-1)\,.
\end{align}
We then
take all explicit momentum
variables within the equations for the Wigner transformed quantities
as conformal. The relation to a physical momentum is given
by $k_{\rm ph}=k/a(\eta)$ and
the time coordinate is understood as the conformal time
$\eta$. Likewise, $T$ denotes a comoving temperature that is related to the
physical temperature as $T_{\rm ph}=T/a(\eta)$, and $\beta=1/T$. The
masses of the right-handed neutrinos $M_i$ are the physical masses. When they occur explicitly
in the Wigner transformed equations, they are accompanied by the scale factor $a(\eta)$
to give the conformally rescaled mass $a(\eta) M_i$.

The Hermitian and anti-Hermitian parts of the
Kadanoff-Baym equations~(\ref{KB:gradexp}) lead us to the constraint
and the kinetic equations
\begin{subequations}
\label{eq:constraint:kin}
\begin{align}
\label{eq:constraint}
2 k^0 {\rm i}\gamma^0 S^{<,>}_\ell
-\left\{
\mathbf k\cdot{\bm \gamma}\gamma^0
+{\Sigma\!\!\!\!/}^H_\ell\gamma^0,{\rm i}\gamma^0 S^{<,>}_\ell
\right\}
-\left\{{\rm i}{\Sigma\!\!\!\!/}^{<,>}_\ell\gamma^0, \gamma^0 S^H_\ell\right\}
&=-\frac12\left({\rm i}{\cal C}_\ell-{\rm i}{\cal C}_\ell^\dagger\right)\,,
\\
\label{eq:kinetic}
{\rm i}\partial_\eta{\rm i}\gamma^0 S^{<,>}_\ell
-\left[
\mathbf k\cdot{\bm \gamma}\gamma^0
+{\Sigma\!\!\!\!/}^H_\ell\gamma^0,{\rm i}\gamma^0 S^{<,>}_\ell
\right]
-\left[{\rm i}{\Sigma\!\!\!\!/}^{<,>}_\ell\gamma^0, \gamma^0 S^H_\ell\right]
&=-\frac12\left({\rm i}{\cal C}_\ell+{\rm i}{\cal C}_\ell^\dagger\right)
\,,
\end{align}
\end{subequations}
with the collision term
\begin{align}
\label{collision:term}
{\cal C}_\ell & = {\rm i}{\Sigma\!\!\!\!/}^>_\ell  
{\rm i}S^<_\ell - {\rm i}{\Sigma\!\!\!\!/}^<_\ell {\rm i}S^>_\ell \,. 
\end{align}

\subsection{Thermal Self Energies}

We now specify the form of the Hermitian self energy 
${\Sigma\!\!\!\!/}^H_\ell$,
which determines the thermal corrections to the dispersion relation. 
For the purpose of this discussion, we assume that the deviation 
of the right handed neutrino distribution from thermal equilibrium 
is small, so that the self energy, being proportional to  
coupling constants, can be evaluated with thermal propagators to 
first order in the gradient expansion. The self energy receives
contributions from the interactions specified
in Eq.~(\ref{TheLagrangian}), but also from 
${\rm SU}(2)_{\rm L}\times{\rm U}(1)_Y$ gauge interactions.
We parametrise this as
\begin{align}
\label{thermal:mass}
{\Sigma\!\!\!\!/}^H_{\ell}
=P_{\rm R}\left[
\gamma^0 (\bar \varsigma^{\rm bl}+\bar \varsigma^{\rm fl})
+\frac{{\mathbf k}\cdot{\bm \gamma}}{|{\mathbf k}|}
\,\big(\varsigma^{\rm bl}+\varsigma^{\rm fl}
-{\rm sign}(k^0)\,[\bar \varsigma^{\rm bl}+\bar \varsigma^{\rm fl}]\big)
\right]
P_{\rm L}
\,,
\end{align}
where we have decomposed the contributions to the self energy into a flavour 
blind part that originates from
${\rm SU}(2)_{\rm L}\times {\rm U}(1)_Y$ gauge interactions
\begin{align}
\label{varsigma:nocoup}
\bar\varsigma^{\rm bl}_{ab}(k^0,\mathbf{k})&=
\delta_{ab} \bar\varsigma^{\rm bl}(k^0,\mathbf{k})\,,
\qquad
\varsigma^{\rm bl}_{ab}(k^0,\mathbf{k})=
\delta_{ab} \varsigma^{\rm bl}(k^0,\mathbf{k})\,,
\end{align}
and a flavour dependent part 
that receives contributions from the charged lepton and the singlet
neutrino Yukawa couplings. To one loop order,
these can be parametrised as
\begin{eqnarray}
\label{varsigmafl:decomp}
&& \bar\varsigma^{\rm fl}_{ab}(k^0,\mathbf{k}) = h^\dagger_{ac} h_{cb} 
\bar\varsigma^{{\rm fl},h}(k^0,\mathbf{k}) +
\sum\limits_i Y^*_{ia}Y_{ib} \bar\varsigma^{{\rm fl},Y}_i(k^0,\mathbf{k})\,, 
\\
&&\varsigma^{\rm fl}_{ab}(k^0,\mathbf{k})=  h^\dagger_{ac} h_{cb} 
\varsigma^{{\rm fl},h}(k^0,\mathbf{k})
+\sum\limits_i Y^*_{ia}Y_{ib} \varsigma^{{\rm fl},Y}_i(k^0,\mathbf{k}) \,.
\nonumber
\end{eqnarray}
In the hierarchical mass limit ($M_1\ll M_{2}$), which we assume within 
this paper, we may restrict the sum to $i=1$. Note that
$\gamma^0{\Sigma\!\!\!\!/}^H_\ell$ is Hermitian, such that
$\varsigma^{\rm fl}$ and $\bar \varsigma^{\rm fl}$ 
are Hermitian matrices in flavour space.

The fact that two independent functions $\varsigma(k^0,\mathbf{k})$ and
$\bar\varsigma(k^0,\mathbf{k})$ occur is because the self energy
${\Sigma\!\!\!\!/}^H_{\ell}$ acquires
contributions that are proportional to $k\!\!\!/$ and $u\!\!\!/$, where
$u^\mu=(1,0,0,0)^T$ is the plasma vector~\cite{Weldon:1982bn}. The relation
between  $\varsigma(k^0,\mathbf{k})$,
$\bar\varsigma(k^0,\mathbf{k})$ and the terms proportional
to $k\!\!\!/$ and $u\!\!\!/$ can be easily established, see {\it e.g.}
Eqs.~(\ref{relation:sigma:ab}). The motivation for our parametrisation is
that it corresponds to a decomposition into a correction $\varsigma$ for the
dispersion relation and a  correction $\bar\varsigma$ that leaves the
dispersion relation unaltered  [see Eqs.~(\ref{kin:eq:g})
and~(\ref{eq:poles}) below]. Besides, $\varsigma$ and $\bar\varsigma$
exhibit useful symmetry properties under the exchange $k^0\to-k^0$
[see Eq.~(\ref{symm:varsigma}) below].

% 
% {\tt Old diagonal form of the self energy }
% 
% where we have introduced in the flavour-diagonal basis the matrices
% \begin{align}
% \bar\varsigma^{\rm bl}_{ab}&=\delta_{ab} \bar\varsigma^{\rm bl}\,,
% \qquad
% \varsigma^{\rm bl}_{ab}=\delta_{ab} \varsigma^{\rm bl}\,,
% \\\notag
% \bar\varsigma^{\rm fl}_{ab}&=\delta_{ab} \bar\varsigma^{\rm fl}_a\,,
% \qquad
% \varsigma^{\rm fl}_{ab}=\delta_{ab} \varsigma^{\rm fl}_a\,.
% \end{align}
% 
% {\tt end of old definition }

The matrix $\varsigma^{\rm fl}$
can be diagonalised through a unitary transformation $U$ and we define
\begin{align}
\varsigma^{\rm fl}_{\rm D} = U^\dagger \varsigma^{\rm fl} U\,.
\end{align}
Note that in general, $U$ is momentum- and time-dependent.
At temperatures $M_{2,3}\gg T/a(\eta)\gg M_1$ and momenta
$|\mathbf k|\sim T$, both  $\varsigma^{{\rm fl},h}$ and
$\varsigma^{{\rm fl},Y}$ are approximately proportional to $T^2/|\mathbf{k}|$, such that the diagonalisation
matrix $U$ is constant in time. When the temperature $T/a(\eta)$
drops below $M_1$, the distribution of $N_1$ becomes Maxwell suppressed.
We consider this situation in Appendix~\ref{appendix:thmass}.
The function $\varsigma_i^{\rm fl,Y}$ then falls
toward zero as $\sim [T/(a M_1)]^4$, {\it cf.} Eq.~(\ref{disp:largeM1}),
and $U$ may generically undergo a change.
Afterwards, at temperatures $T/a(\eta)\ll M_1$, $U$ becomes constant again
and the matrix $U^\dagger h^\dagger h U$ is diagonal.

All quantities that carry left-handed flavour indices transform under the 
basis transformation defined by $U$. We denote matrices evaluated in the 
flavour-diagonal basis by a subscript ${\rm D}$. For example,
\begin{subequations}
\begin{align}
\label{sig:diag}
{\Sigma\!\!\!\!/}^H_{\ell \rm D}  &= U^\dagger {\Sigma\!\!\!\!/}^H_{\ell} U\,,
\\
\label{S:diag}
{\rm i}S_{\ell \rm D}^{fg} &= U^\dagger {\rm i} S_{\ell }^{fg} U\,.
\end{align}
\end{subequations}
Note that unlike $\varsigma^{\rm fl}_{\rm D}$, these matrices are in 
general not diagonal in flavour-space.

Inserting the definitions~(\ref{sig:diag}) and~(\ref{S:diag}) into~(\ref{eq:kinetic}) and multiplying with $U^\dagger$ from the left and with $U$ from the right, we obtain the kinetic equation in the lepton thermal mass basis\footnote{The terms
${\bm \gamma}\gamma^0$ and $\gamma^0 S^{<,>}_\ell$ commute when using the 
ansatz~(\ref{S:decomposition}) below.
}
\begin{align}
\label{kin:eq:diag}
   {\rm i} \partial_\eta {\rm i} \gamma^0 S^{<,>}_{\ell \rm D} +{\rm i} \left[ \Xi , {\rm i} \gamma^0 S_{\ell \rm D}^{<,>} \right] - \left[  {\Sigma\!\!\!\!/}^H_{\ell \rm D}\gamma^0, {\rm i} \gamma^0 S_{\ell \rm D}^{<,>} \right] & = \frac{1}{2}\left( {\rm i}{\cal C}_{\ell \rm D} + {\rm i}{\cal C}_{\ell \rm D}^\dagger \right)\,,
\end{align}
where ${\cal C}_{\ell \rm D} = U^\dagger {\cal C}_\ell U$ and
\begin{align}
\Xi = U^\dagger \partial_\eta U
\end{align}
is the compensation matrix for time-dependent flavour rotations.
At first order in gradients, the only consequence of a time dependent $U$ is the additional term involving $\Xi$. Therefore, we switch to the diagonal basis and drop the subscript ${\rm D}$ from all subsequent expressions.

\subsection{Kinetic and Constraint Equations}
\label{section:massshells}

The Weyl fermion propagator can be parametrised through the vector and
pseudovector functions
\begin{align}
\label{S:decomposition}
{\rm i}\gamma^0 S_\ell^{<,>}=
\frac 12 \sum\limits_{h=\pm}
\left[
g_{h0}^{<,>}
\left(
\mathbbm 1 + h \hat{\mathbf k}\cdot \gamma^5\gamma^0{\bm \gamma}
\right)
+g_{h3}^{<,>}
\left(
\gamma^5 + h\hat{\mathbf k}\cdot \gamma^0{\bm \gamma}
\right)
\right]\,,
\end{align}
where $\hat{\mathbf k} ={\mathbf k}/|{\mathbf k}|$.
When compared to the case of a Dirac fermion,
there are no scalar and pseudoscalar contributions. This is because gauge symmetry 
prevents the dynamical generation of scalar, pseudoscalar and tensor 
densities, provided the gauge symmetry is neither broken spontaneously 
nor through initial conditions, as we assume here. Thus, 
Eq.~(\ref{S:decomposition}) is the most general form of the lepton 
propagator compatible with isotropy and chiral symmetry. Besides, 
from the fact that the leptons $\ell$ are left-handed, we immediately 
obtain
\begin{align}
\label{Weyl:constraint}
g_{h0}^{<,>}=g_{h3}^{<,>}\equiv g_h^{<,>}\,.
\end{align}
We furthermore see that
$\mathbf{k}\cdot {\bm \gamma} \gamma^0$ and ${\rm i}\gamma^0S_\ell^{<,>}$
commute, such that the constraint and kinetic Eqs.~(\ref{eq:constraint:kin})
simplify to
\begin{subequations}
\begin{align}
\label{eq:constraint:comm}
2 (k^0-\mathbf k\cdot{\bm \gamma}\gamma^0) {\rm i}\gamma^0 S^{<,>}_\ell
-\left\{{\Sigma\!\!\!\!/}^H_\ell\gamma^0,{\rm i}\gamma^0 S^{<,>}_\ell\right\} -\left\{{\rm i}{\Sigma\!\!\!\!/}^{<,>}_\ell\gamma^0, \gamma^0 S^H_\ell\right\}
&=-\frac12\left({\rm i}{\cal C}_\ell-{\rm i}{\cal C}_\ell^\dagger\right)\,,
\\
\label{eq:kinetic:comm}
{\rm i}\partial_\eta{\rm i}\gamma^0 S^{<,>}_\ell
-\left[{\Sigma\!\!\!\!/}^H_\ell\gamma^0,{\rm i}\gamma^0 S^{<,>}_\ell\right] -\left[{\rm i}{\Sigma\!\!\!\!/}^{<,>}_\ell\gamma^0, \gamma^0 S^H_\ell\right]
&=-\frac12\left({\rm i}{\cal C}_\ell+{\rm i}{\cal C}_\ell^\dagger\right)
\,.
\end{align}
\end{subequations}

To zeroth order,
the constraint equation~(\ref{eq:constraint}) reduces to the simple form
\begin{align}
\label{eq:constraint:eq}
\left\{
k\!\!\!/,
{\rm i} S^{<,>}_\ell
\right\}
=0\,.
\end{align}
Substituting the ansatz~(\ref{S:decomposition})
leads us to
\begin{subequations}
\begin{align}
g_{h0}^{<,>} k^0 + h |\mathbf k| g_{h3}^{<,>}
&=0\,,
\\
g_{h3}^{<,>} k^0 + h |\mathbf k| g_{h0}^{<,>}
&=0
\,.
\end{align}
\end{subequations}
The constraint~(\ref{Weyl:constraint}) then implies that 
$g_{h}^{<,>}(k^0,{\mathbf k})$ is non-vanishing only when 
$k^0 = -h |{\mathbf k}|$, which corresponds to the singular 
zero-mass shell. In particular $h=-\mbox{sign}(k^0)$, 
that is for leptons ($k^0>0$) the helicity $h=-1$ is negative, while for
anti-leptons ($k^0<0$) the helicity $h=1$ is positive, as 
expected. This relation is weakly
broken for momenta $|\mathbf{k}|\sim T$
when including thermal corrections, because hole modes exhibit an opposite
connection between frequency and helicity. For momenta $|\mathbf{k}|\ll T$,
the hole modes couple to the plasma at a similar strength as the particle 
modes. However, this region only corresponds to a small portion of the 
available phase space, such that we may neglect it here.

Substituting the parametrisations (\ref{thermal:mass}),~(\ref{S:decomposition})
and the constraint~(\ref{Weyl:constraint}) into the kinetic
equations~(\ref{eq:kinetic:comm}) and taking the trace leads us to
\begin{align}
\label{kin:eq:g}
{\rm i} \partial_\eta g_{h}^{<,>} + {\rm i} \left[ \Xi, g_{h}^{<,>} \right] + h \left[ \varsigma^{\rm fl},g_{h}^{<,>} \right] = -\frac{1}{4} {\rm tr}\left({\rm i}{\cal C}_\ell + {\rm i} {\cal C}_\ell^\dagger  \right),
\end{align}
where the trace is taken only in Dirac space, and $\varsigma^{\rm fl}$ and all other objects are evaluated in the mass-diagonal basis. 
As explained above, the
helicity is determined from the zeroth order
constraint equation by the relation $h = - {\rm sign}(k^0)$. 
Eq.~(\ref{kin:eq:g}) is accurate up to first order in gradients, because
$\varsigma^{\rm fl}$ itself is of first order. 
We note that the term 
$\left[{\rm i}{\Sigma\!\!\!\!/}^{<,>}_\ell\gamma^0, \gamma^0 S^H_\ell\right]$ 
in Eq.~(\ref{eq:kinetic:comm}) does not contribute to 
Eq.~(\ref{kin:eq:g}) at first order, since 
first, $S^H_\ell$ can be evaluated at zeroth order, where 
is it independent 
of the particle distribution functions and therefore proportional to 
the unit matrix in flavour space; and second, in Dirac space the 
commutator reads
\begin{equation}
\left[{\rm i}{\Sigma\!\!\!\!/}^{<,>}_\ell\gamma^0, \gamma^0 S^H_\ell\right] 
\propto - {\rm i} 
\big[P_{\rm R}\,(a^{<,>} \not\!k+b^{<,>}\gamma^0)P_{\rm L} \gamma^0,
\gamma^0P_{\rm L}\!\not\!kP_{\rm R}\big] =0\,.
\end{equation}
Since we work in the flavour-diagonal basis, where $\varsigma^{\rm fl}$ 
is diagonal, the commutator involving $\varsigma^{\rm fl}$ in 
Eq.~(\ref{kin:eq:g}) can be explicitly evaluated, which yields
\begin{align}
\label{kinetic:equation}
{\rm i}\partial_\eta g^{<,>}_{hab}
+ {\rm i} \left[ \Xi, g_{h}^{<,>} \right]_{ab}
+h(\varsigma^{\rm fl}_{aa}-\varsigma^{\rm fl}_{bb})g^{<,>}_{hab}
=-\frac14
\left(
{\rm tr}\left[{\rm i}{\cal C}_\ell+{\rm i}{\cal C}_\ell^\dagger\right]
\right)_{ab}
\,.
\end{align}
Hence, the thermal dispersion relations have the same impact on the equation
of motion for the lepton density as explicit Dirac masses
would~\cite{Konstandin:2004gy,Konstandin:2005cd,Cirigliano:2009yt}. 

An important feature of Eq.~(\ref{kin:eq:g})
is the sign change of the commutator term
involving the thermal dispersion relation through
$\varsigma^{\rm fl}$ when $h\to-h$ or, alternatively, $k^0\to - k^0$.
%This finds an intuitive explanation when noting that at one loop
%order, the thermal
%dispersion relation for leptons
%must not exhibit a strong $CP$-phase (a phase that originates
%from the presence of on-shell intermediate states and that
%adds with the same sign to a Feynman diagram and its $CP$ conjugate,
%while a weak phase adds with opposite sign).
We now show that at the one-loop level, the elements of $h\varsigma^{\rm fl} = 
- \mbox{sign}(k^0) \,\varsigma^{\rm fl}(k^0,\mathbf{k})$ are indeed 
odd functions in $k^0$.
%in the limit of
%absence of strong $CP$-phases.
We define charge and parity conjugation through
\begin{align}
\psi^C(x)&=C\bar\psi^T(x)\,,\\
\psi^P(x)&=P\psi(\bar x)\,,
\end{align}
where $\bar u \equiv (u_0, -\mathbf{u})$ and where
in the Weyl representation, the conjugation 
matrices are given by $C = i\gamma^0\gamma^2$ and $P = \gamma^0$.
Thereby, we fix possible $CP$ phases that can arise in the definition
of these conjugations to zero.
The charge and parity conjugate propagators are
\begin{subequations}
\label{chargeparity:conj}
\begin{align}
\label{charge:conj}
{\rm i}S^{C,fg}_{\ell ab}(u,v)
=&\langle \psi^C_{\ell a}(u^f) \bar{\psi}^C_{\ell b}(v^g)\rangle
=C \left[{\rm i}S^{gf}_{\ell ba} (v,u)\right]^{T} C^\dagger\,,
\\
\label{parity:conj}
{\rm i}S^{P,fg}_{\ell ab}(u,v)
=&\langle \psi^P_{\ell a}(u^f) \bar{\psi}^P_{\ell b}(v^g)\rangle
=P {\rm i} S^{fg}_{\ell ab} (\bar u, \bar v) P^\dagger\,.
\end{align}
\end{subequations}
The transposition acts here only on the Dirac indices,
which in contrast to the flavour and CTP indices are not written explicitly. 
The $CP$ conjugate of the Hermitian self energy is then given by
(we suppress the average coordinate in the argument of $\varsigma$
and  $\bar\varsigma$)
\begin{align}
\label{SigmaH:conjugate}
%{\Sigma\!\!\!\!/}^{H*}_{\ell ab}(k,x)
%&=
{\Sigma\!\!\!\!/}^{CP, H}_{\ell ab}(k,x)
&=
CP \left[{\Sigma\!\!\!\!/}^{H}_{\ell ba} (- \bar k, \bar x)\right]^{T} (CP)^\dagger
% \\
% \notag
% &=
% PC\Big\{
% P_{\rm R}
% \Big[
% \gamma^0 (\bar \varsigma_{ba}^{\rm bl}(-\bar k)+\bar \varsigma_{ba}^{\rm fl}(-\bar k))
% +\hat {\mathbf k}\cdot{\bm \gamma}
% (\varsigma_{ba}^{\rm bl}(-\bar k)+\varsigma_{ba}^{\rm fl}(-\bar k)
% \\
% \notag
% &\hskip1.4cm
% +{\rm sign}(k^0)[\bar \varsigma_{ba}^{\rm bl}(-\bar k)+\bar \varsigma_{ba}^{\rm fl}(-\bar k)])
% \Big]
% P_{\rm L}
% \Big\}^T C^\dagger P^\dagger
% \\
% \notag
% &=
% P\Big\{
% P_{\rm L}
% \Big[
% -\gamma^0 (\bar \varsigma_{ba}^{\rm bl}(-\bar k)+\bar \varsigma_{ba}^{\rm fl}(-\bar k))
% -\hat {\mathbf k}\cdot{\bm \gamma}
% (\varsigma_{ba}^{\rm bl}(-\bar k)+\varsigma_{ba}^{\rm fl}(-\bar k)
% \\
% \notag
% &\hskip1.4cm
% +{\rm sign}(k^0)[\bar \varsigma_{ba}^{\rm bl}(-\bar k)+\bar \varsigma_{ba}^{\rm fl}(-\bar k)])
% \Big]
% P_{\rm R}
% \Big\} P^\dagger
\\
\notag
&=
P_{\rm R}
\Big[
-\gamma^0 (\bar \varsigma_{ba}^{\rm bl}(-\bar k)+\bar \varsigma_{ba}^{\rm fl}(-\bar k))
+\hat {\mathbf k}\cdot{\bm \gamma}
(\varsigma_{ba}^{\rm bl}(-\bar k)+\varsigma_{ba}^{\rm fl}(-\bar k)
\\
\notag
&\hskip1.4cm
+{\rm sign}(k^0)[\bar \varsigma_{ba}^{\rm bl}(-\bar k)+\bar \varsigma_{ba}^{\rm fl}(-\bar k)])
\Big]
P_{\rm L}
\,.
\end{align}
On the other hand, we may calculate this self energy from the
$CP$-conjugate Lagrangian, within which the coupling constants are
complex conjugated, as
\begin{align}
{\Sigma\!\!\!\!/}^{CP,H}_{\ell ab}(k,x)
={\Sigma\!\!\!\!/}^{H}_{\ell ab}(k,x)\Big|_{
\stackrel{Y\to Y^*}{h\to h^*}
}\,,
\end{align}
provided the initial conditions preserve $CP$ symmetry (no primordial 
asymmetry).
To the one-loop order, the coupling constants appear as the prefactors
$h^\dagger  h$ and $Y^\dagger Y$
within ${\Sigma\!\!\!\!/}^{H}_{\ell}$, {\it cf.} Eq.~(\ref{varsigmafl:decomp}).
The effect of $CP$ conjugation
therefore amounts to the replacements
$[h^\dagger  h]_{ab} \to [h^\dagger h]^*_{ab}=[h^\dagger h]_{ba}$ and
$[Y^\dagger Y]_{ab} \to [Y^\dagger Y]^*_{ab}=[Y^\dagger Y]_{ba}$, and it
follows that to one-loop order
\begin{align}
{\Sigma\!\!\!\!/}^{CP,H}_{\ell ab}(k,x)
={\Sigma\!\!\!\!/}^{H}_{\ell ba}(k,x)\,.
\end{align}
Comparing this to the relation~(\ref{SigmaH:conjugate}) and substituting
Eq.~(\ref{thermal:mass}),
we find that 
\begin{align}
\label{symm:varsigma}
\bar\varsigma^{\rm fl}_{ab}(-k^0,\mathbf{k}) = -\bar\varsigma^{\rm fl}_{ab}(k^0,\mathbf {k}) \qquad {\rm and} \qquad \varsigma^{\rm fl}_{ab}(-k^0,\mathbf{k}) = \varsigma^{\rm fl}_{ab}(k^0,\mathbf{k})\,,
\end{align}
and accordingly for $\bar\varsigma^{\rm bl}$ and $\varsigma^{\rm bl}$, 
which implies the result to be shown for the elements of 
$\varsigma^{\rm fl}$. These symmetry properties with
respect to $k^0$ are in accordance with the equilibrium results from
Ref.~\cite{Weldon:1982bn} and Appendix~\ref{appendix:thmass}. The present 
argument shows moreover that
they also hold under out-of-equilibrium conditions.

For calculating the momentum integrals in the collision terms, we
use on-shell conditions that we obtain from the constraint equations.
In order to achieve accuracy to first order in gradients, it again suffices to
solve the constraint equations to zeroth order,
since the collision term is suppressed by higher orders in the
coupling constants.
A similar approximation is applied in Ref.~\cite{Cirigliano:2009yt},
where in contrast to the present work, within the constraint equations,
small tree-level mass differences rather than differences between 
one-loop dispersion relations and finite widths are neglected.
A general solution to the zeroth-order
constraint equation~(\ref{eq:constraint:eq})
is given by the Kadanoff-Baym ansatz
\begin{subequations}
\label{Slessgreater:ansatz}
\begin{align}
\label{Seq:less}
{\rm i}S^<_{\ell ab}&=-2 S^{\cal A}_{\ell}
\left[
\vartheta(k^0)f^{+}_{\ell ab}(\mathbf k)
-\vartheta(-k^0)(\mathbbm{1}_{ab}-f^{-}_{\ell ab}(-\mathbf k))
\right]\,,
\\
\label{Seq:greater}
{\rm i}S^>_{\ell ab}&=-2 S^{\cal A}_{\ell}
\left[
-\vartheta(k^0)(\mathbbm{1}_{ab}-f^{+}_{\ell ab}(\mathbf k))
+\vartheta(-k^0) f^{-}_{\ell ab}(-\mathbf k)
\right]\,.
\end{align}
\end{subequations}
where
\begin{align}
\label{S^A:singular}
S_\ell^{\cal A}=
\pi
P_{\rm L} k\!\!\!/ P_{\rm R}
 \delta \!\left(k^2\right)
\,,
\end{align}
and $f^\pm_{\ell ab}$ are the distribution function matrices of leptons and 
anti-leptons. Comparing Eqs.~(\ref{Slessgreater:ansatz}) to 
Eq.~(\ref{S:decomposition}) we identify
\begin{eqnarray}
&&g^{<}_{-} = -2\pi \delta(k^2)\vartheta(k^0) |\mathbf{k}| f^+_\ell(\mathbf{k})\,,
\\
&&g^{>}_{-} = 2\pi \delta(k^2)\vartheta(k^0) |\mathbf{k}| 
(\mathbbm{1}-f^+_\ell(\mathbf{k}))\,,
\nonumber\\
&&g^{<}_{+} = -2\pi \delta(k^2)\vartheta(-k^0) |\mathbf{k}| 
(\mathbbm{1}-f^-_\ell(-\mathbf{k}))\,,
\nonumber\\
&&g^{>}_{+} = 2\pi \delta(k^2)\vartheta(-k^0) |\mathbf{k}| 
f^-_\ell(-\mathbf{k})\,,
\nonumber
\end{eqnarray}
which is
consistent with $h=-\mbox{sign}(k^0)$ and $k^0=\pm|\mathbf{k}|$.
While for the present work, these zeroth-order solutions to the constraint 
equations are sufficient, we show in Appendix~\ref{appendix:finitewidth} 
how to extend them to first order in consistency
with the equations for the retarded and the advanced 
propagators~(\ref{polemass:gradexp}).

Note that if we identify $f^\pm_{\ell ab}$ with expectation values of 
number density operators then it follows from 
Eqs.~(\ref{Slessgreater:ansatz}) and the operator definition of 
${\rm i}S^{<,>}_{\ell ab}$ that $f^+_{\ell ab} \sim 
\langle a_b^\dagger a_a\rangle$ corresponds to the lepton density 
matrix, while  $f^-_{\ell ab} \sim 
\langle b_a^\dagger b_b\rangle$ corresponds to the {\em transpose} 
of the anti-lepton densities. This may also be seen when relating
the $CP$ conjugate to the original propagator as
\begin{align}
\label{S:CP}
{\rm i}S^{CP,fg}_{\ell ab}(k,x)
&=CP\left[{\rm i}S^{gf}_{\ell ba}(-\bar k,\bar x)\right]^T(CP)^\dagger
\\
\notag
&=
-\frac12\sum\limits_{h=\pm}
g^{gf}_{hba}(- k)
\left[
(1-\gamma^5)\gamma^0-(1-\gamma^5)h \hat{\mathbf{k}} \cdot {\bm \gamma}
\right]
=-{\rm i}S^{gf}_{\ell ba} (-k,x)\,,
\end{align}
where we have used the ansatz~(\ref{S:decomposition})
and have assumed spatial homogeneity, $g^{gf}_h(k)=g^{gf}_h(\bar k)$.
Therefore, a sign flip in $k$ yields the negative of the 
flavour- and CTP-transposed $CP$ conjugate propagator.
The use of $S^{<,>}_{\ell} (k,x)$ with $k^0<0$ rather than
$S^{CP,<,>}_{\ell}(k,x)$ with $k^0>0$ to describe the
anti-lepton densities has the advantage that the resulting
kinetic equations are flavour covariant.
This is because the former propagator has the same flavour transformation
properties as $S^{<,>}_{\ell} (k,x)$ with $k^0>0$,
while the latter transforms in
the complex conjugate representation. This property
has been described and used before in the context of electroweak
baryogenesis~\cite{Konstandin:2004gy,Konstandin:2005cd}.

\section{Kinetic Equations for Lepton Number Densities}
\label{sec:kinetic}

In this Section, we perform simplifications of the kinetic
equations~(\ref{kinetic:equation}), such that they attain a form which can be
solved numerically. The key simplification arises from
the separation of the time scales of kinetic equilibration and 
flavour-sensitive interactions. Because the former is much faster,
the distribution functions are driven to kinetic equilibrium, such
that they can be approximated by the Bose-Einstein or 
Fermi-Dirac form, parametrised through a matrix of chemical potentials.
An integration over the momentum then allows to express the kinetic equations
in terms of charge densities. We shall often use spatial homogeneity
to set $f^\pm_{\ell ab}(-\mathbf k) = f^\pm_{\ell ab}(\mathbf k)$.

\subsection{Matrices for Lepton Number Densities}
The lepton number density matrices are defined as
\begin{subequations}
\label{relate:nu:S}
\begin{align}
n_{\ell ab}^{+}=
\int\frac{d^3 k}{(2\pi)^3}
f^+_{\ell ab}(\mathbf k)
&=-
\int\frac{d^3 k}{(2\pi)^3}
\int_{0}^{\infty}
\frac{d k^0}{2\pi}
{\rm tr}\left[
{\rm i}\gamma^0 S_{\ell ab}^{<}
\right] \,,
\\
n_{\ell ab}^{-}=
\int\frac{d^3 k}{(2\pi)^3}
f^-_{\ell ab}(\mathbf k)
&=
\int\frac{d^3 k}{(2\pi)^3}
\int_{-\infty}^{0}
\frac{d k^0}{2\pi}
{\rm tr}\left[
{\rm i}\gamma^0 S_{\ell ab}^{>}
\right] \,.
\end{align}
\end{subequations}
Due to the presence of the fast gauge interactions, we may 
assume kinetic equilibrium for the leptons, and denote
$\delta n_{\ell ab}^\pm=n_{\ell ab}^\pm-n_{\ell ab}^{\pm{\rm eq}}$
and $\delta f_{\ell ab}^\pm=f_{\ell ab}^\pm-f_{\ell ab}^{\pm{\rm eq}}$.
Then, we can introduce a matrix of generalised chemical potentials 
$\mu^\pm_{ab}$ for particles and antiparticles, and write 
\begin{align}
\label{general:chem}
f^\pm_{\ell ab}(\mathbf{k})=
\left(\frac{1}
{
{\rm e}^{\beta |\mathbf{k}|-\beta\mu^\pm}+1
}\right)_{ab}
\,,
\end{align}
such that the number density matrices are, to first order in 
the chemical potentials, 
\begin{align}
\delta n^\pm_{\ell ab}=\mu^\pm_{ab}\frac{T^2}{12}\,.
\end{align}
This allows to relate the lepton number densities to the
distribution functions:
\begin{align}
\label{delta:f}
\delta f^{\pm}_{\ell ab}(\mathbf k)=12\delta n^\pm_{\ell ab}
\frac{\beta^3{\rm e}^{\beta |\mathbf k|}}{({\rm e}^{\beta |\mathbf k|}+1)^2}
\,.
\end{align}
We also introduce the deviation of the $<,>$ propagators from equilibrium,
\begin{align}
\label{delta:S}
{\rm i}\delta S_{\ell ab}
=-2S_\ell^{\cal A}\,[
\vartheta(k^0)\delta f^+_{ab}(\mathbf{k})
+\vartheta(-k^0)\delta f^-_{ab}(-\mathbf{k})
]
\,.
\end{align}
Besides, we use corresponding approximations and expressions for the 
right-handed leptons of the Standard Model by replacing $\ell \to {\rm R}$.

The ansatz~(\ref{general:chem}) is valid provided the interactions
that establish kinetic equilibrium, which are the pair creation and 
annihilation processes and scatterings with gauge bosons, are faster 
than processes that distinguish between the particle flavours, in 
particular flavour oscillations and flavour-sensitive damping rates.
Provided these assumptions hold, which is verified in 
Section~\ref{sec:flblind},
the time scales for kinetic equilibration and for flavour effects separate.
Since $\mu^\pm$ is Hermitian, it is always possible to bring
Eq.~(\ref{general:chem}) to diagonal form by a flavour rotation. Due
to the separation of time scales, kinetic equilibrium
in this diagonal basis is then attained at a rate that is faster
than the flavour effects
that may change the flavour orientation of $\mu_{ab}$.

Using the ansatz~(\ref{Slessgreater:ansatz}) and the spectral
function~(\ref{S^A:singular}) together with the 
decomposition~(\ref{S:decomposition})
and the constraint~(\ref{Weyl:constraint}) gives the relations
\begin{align}
f_\ell^\pm(\mathbf{k})=
\mp 2 \int\limits_{0,-\infty}^{\infty,0} 
\frac{d k^0}{2\pi}g_{\mp}^{<,>}(k^0,\mathbf{k})\,.
\end{align}
Performing a
$k^0$-integration of Eq.~(\ref{kinetic:equation}) and using
Eqs.~(\ref{relate:nu:S}) then yields
\begin{align}
\label{kin:eq:f}
\partial_\eta f_{\ell ab}^\pm(\mathbf k)
=[\Xi, f^\pm_{\ell}(\mathbf k)]_{ab}
\mp {\rm i}(\varsigma^{\rm fl}_{aa} -\varsigma^{\rm fl}_{bb})
f_{\ell ab}^\pm(\mathbf{k})
\pm\frac12\,
{\rm tr}\int\limits_{0,-\infty}^{\infty,0}
\frac{d k^0}{2\pi}
 ({\cal C}_{\ell ab}+{\cal C}^\dagger_{\ell ab})\,.
\end{align}
Integrating over three momenta and applying the expansions explained above,
we find
\begin{align}
\label{kin:eq:nu}
\partial_\eta \delta n_{\ell ab}^\pm
= [\Xi^{\rm eff},\delta n^\pm_\ell ]_{ab}
\mp{\rm i}\Delta\omega_{\ell ab}^{\rm eff}  \delta n_{\ell ab}^\pm
\pm\frac12\,{\rm tr}\int\limits_{0,-\infty}^{\infty,0}\frac{dk_0}{2\pi}
\int\frac{d^3 k}{(2\pi)^3}({\cal C}_{\ell ab}+{\cal C}^\dagger_{\ell ab})\,,
\end{align}
where we have used that the equilibrium distributions are diagonal in flavour, 
$n_{\ell ab}^{\pm{\rm eq}} = \delta _{ab}n_{\ell aa}^{\pm{\rm eq}}$.
We have defined here the thermally averaged frequencies of
flavour oscillations and the compensation matrix as
\begin{subequations}
\begin{align}
\Delta\omega_{\ell ab}^{\rm eff}(\eta)
&=\int\frac{d^3 k}{(2\pi)^3}\,
\frac{12 \beta^3 {\rm e}^{\beta |\mathbf k|}}{({\rm e}^{\beta |\mathbf
    k|}+1)^2}\,
(\varsigma^{\rm fl}_{aa}(|\mathbf{k}|,\mathbf{k},\eta)-
\varsigma^{\rm fl}_{bb}(|\mathbf{k}|,\mathbf{k},\eta))
\,,\\
\Xi^{\rm eff}(\eta)
&=\int\frac{d^3 k}{(2\pi)^3}\,
\frac{12 \beta^3 {\rm e}^{\beta |\mathbf k|}}{({\rm e}^{\beta |\mathbf
    k|}+1)^2}\,
\Xi(|\mathbf{k}|,\mathbf{k},\eta)\,,
\end{align}
\end{subequations}
and used that $\varsigma^{\rm fl}$ and $\Sigma$ are symmetric in their
first argument, {\it cf.} (\ref{symm:varsigma}).
The dominant contributions to the phase space integrals originate from
regions where $|\mathbf k|\sim T$, where $\varsigma_{ab}^{\rm fl}$
can be approximated by~\cite{Weldon:1982bn}
\begin{align}
\label{mell:th}
\varsigma^{\rm
  fl}_{ab}(k^0,\mathbf{k})=\frac{h^{\dagger}_{ac}h_{cb}T^2}
  {16|\mathbf k|}
+\sum\limits_i Y^*_{ia}Y_{ib}\varsigma_i^{{\rm fl},Y}(k^0,\mathbf{k})
\,.
\end{align}
While the form of $Y^\dagger Y\varsigma_i^{{\rm fl},Y}$ is more 
complicated in general, as we discuss in
Appendix~\ref{appendix:thmass}, it is of the same order or smaller
than $h^\dagger h \,\varsigma^{{\rm fl},h}$ when flavour effects are 
important. We can therefore estimate
\begin{align}
\Delta\omega_{\ell ab}^{\rm eff}={\cal O}
\left(
h_\tau^2 T
\right)\,,
\end{align}
where $h_\tau$ is the $\tau$-lepton Yukawa coupling.

We decompose the collision term\footnote{The definition of
the collision terms ${\cal C}$ in Ref.~\cite{Beneke:2010wd}
differs from the present ones by an additional integration $\int dk^0/(2\pi)$.
} as
\begin{align}
{\cal C_\ell}={\cal C}_\ell^Y+{\cal C}_\ell^{\rm fl}+{\cal C}_\ell^{\rm bl}\,.
\end{align}
The term ${\cal C}_\ell^Y$ describes decays and inverse decays of $N_1$ and
hence the washout and the $CP$-asymmetric source of the lepton densities.
Flavour sensitive interactions mediated by the Standard Model Yukawa couplings
are encompassed in ${\cal C}_\ell^{\rm fl}$, while flavour-blind interactions
mediated by gauge couplings are taken account of 
within ${\cal C}_\ell^{\rm bl}$. In the following,
we show that these particular contributions can be cast into the form
\begin{align}
\label{kin:eq_nu}
\frac{\partial \delta n^\pm_{\ell ab}}{\partial \eta}
=&\,\,
\Xi^{\rm eff}_{ac} \delta n_{\ell cb}^\pm -
\delta n_{\ell ac}^\pm\Xi^{\rm eff}_{cb}
\mp{\rm i}\Delta\omega^{\rm eff}_{\ell ab} \delta n^\pm_{\ell ab}
\\
\notag
&- \sum\limits_{c}[W_{ac}\delta n^\pm_{\ell cb}
+\delta n^{\pm *}_{\ell ca}W_{bc}^*]
\pm S_{ab}
-\Gamma^{\rm bl}(\delta n^+_{\ell ab}+\delta n^-_{\ell ab})
-\Gamma_{\ell ab}^{\pm \rm fl}
\,.
\end{align}

\subsection{Source and Washout Term}
The washout term for $\delta n_{\ell ab}^+$ is given by
\begin{align}
-\sum\limits_{c} W_{ac}\delta n^+_{\ell cb}
&=
\frac 12\,{\rm tr}
\int\frac{d^3 k}{(2\pi)^3}\int_0^\infty\frac{d k_0}{2\pi}
\,{\cal C}_{\ell ab}^{Y}
\\\notag
&=
\sum\limits_{c}\int\frac{d^3 k}{(2\pi)^3}\int_0^\infty\frac{d
  k_0}{2\pi}
\,\frac12{\rm tr}
\,\left[
{\rm i}{\Sigma\!\!\!/}^>_{\ell ac}(k)
{\rm i}S^<_{\ell cb}(k)
-
{\rm i}{\Sigma\!\!\!/}^<_{\ell ac}(k)
{\rm i}S^>_{\ell cb}(k)
\right]\,,
\end{align}
evaluated to order $Y_{ia}^2$ (certain higher order terms are accounted
for by the source term).
Close to equilibrium, we can write
\begin{align}
{\rm i}{\Sigma\!\!\!/}^>_{\ell ac}(k)
{\rm i}S^<_{\ell cb}(k)
-
{\rm i}{\Sigma\!\!\!/}^<_{\ell ac}(k)
{\rm i}S^>_{\ell cb}(k)
=-{\rm i}\left({\Sigma\!\!\!/}^<_{\ell ac}(k)-
{\Sigma\!\!\!/}^>_{\ell ac}(k)\right) {\rm i}\delta S_{\ell cb}(k)\,,
\end{align}
and we note that
\begin{align}
{\rm i}{\Sigma\!\!\!/}^<_{\ell ac}(k)-
{\rm i}{\Sigma\!\!\!/}^>_{\ell ac}(k)
=&-Y_{1a}^*Y_{1c}\!\int\!\!\frac{d^3 k^\prime}
{(2\pi)^3 2\sqrt{{\mathbf{k}^\prime}^2+(a(\eta)M_1)^2}}
\frac{d^3 k^{\prime\prime}}{(2\pi)^3 2|\mathbf k^{\prime\prime}|}
\,(2\pi)^4 \delta^4(k^\prime-k-k^{\prime\prime})
\\\notag
&\times
{\rm sign}(k_0)
P_{\rm R}(k\!\!\!/^\prime +a(\eta) M_1)P_{\rm L}
\left(
f_{Ni}(\mathbf k^{\prime})+f_{\phi}(\mathbf k^{\prime\prime})
\right)\,.
\end{align}
Substituting Eqs.~(\ref{delta:f}) and~(\ref{delta:S}),
we can identify
\begin{align}
W_{ac}
=&\frac 12 \,Y_{1a}^* Y_{1c}
\int\!
\frac{d^3 k}{(2\pi)^3 2|\mathbf k|}
\frac{d^3 k^\prime}{(2\pi)^3 2\sqrt{{\mathbf{k}^\prime}^2+(a(\eta)M_1)^2}}
\frac{d^3 k^{\prime\prime}}{(2\pi)^3 2|\mathbf k^{\prime\prime}|}
(2\pi)^4 \delta^4(k^\prime-k-k^{\prime\prime})
\\\notag
&\times
2k\cdot k^\prime
\left(
f_{N1}(\mathbf k^\prime)+f_{\phi}(\mathbf k^{\prime\prime})
\right)\,
\frac{12 \beta^3\,{\rm e}^{\beta |\mathbf k|}}{({\rm e}^{\beta |\mathbf k|}+1)^2}
\,.
\end{align}
The washout term for $\delta n_{\ell ab}^-$ follows correspondingly.

In straightforward generalisation of the single flavour case, the
$CP$-violating source term for $\delta n_{\ell ab}^+$ is
\begin{align}
S_{ab}=&\,\,
\frac32\,{\rm i}\sum\limits_c
\left[
Y_{1a}^*Y_{1c}^*Y_{2c}Y_{2b}
-Y_{2a}^*Y_{2c}^*Y_{1c}Y_{1b}
\right]
\\
\notag
& \times
\left(-\frac{M_1}{M_2}\right)
\int\frac{d^3k^\prime}{(2\pi)^32\sqrt{{\mathbf k^\prime}^2+(a(\eta) M_1)^2}}
\frac{\Sigma_{N\mu}(\mathbf k^\prime) \Sigma_N^\mu(\mathbf k^\prime)}{g_w}
\delta f_{N1}(\mathbf k^\prime)\,,
\end{align}
where~\cite{Beneke:2010wd}
\begin{align}
\Sigma_N^\mu(k)=g_w\int\frac{d^3p}{(2\pi)^3 2|\mathbf p|}
\frac{d^3q}{(2\pi)^3 2|\mathbf q|}
(2\pi)^4 \delta^4(k-p-q)\,p^\mu
\left(
1-f_\ell^{\rm eq}(\mathbf p)+f_\phi^{\rm eq}(\mathbf q)
\right)\,,
\end{align}
with $g_w=2$,
and where the source for the anti-leptons $\delta n_{\ell ab}^-$ is $-S_{ab}$.
The deviation of the distribution function of the right-handed  neutrinos
from equilibrium is denoted by 
$\delta f_{N1}(\mathbf k)= f_{N1}(\mathbf k)- f^{\rm eq}_{N1}(\mathbf k)$.
This source term is understood to include both, wave-function and vertex
contributions in the hierarchical limit, $M_1 \ll M_2$. Note that there
is an additional wave-function contribution that violates lepton flavour
but conserves total lepton number~\cite{Endoh:2003mz,Nardi:2006fx}.
Within flavoured models of leptogenesis,
this may contribute to the final lepton asymmetry. Compared to the
total lepton-number violating contributions, it is however suppressed
by a factor $M_1/M_2$ since one picks up the 
$\not\!k\sim M_1$ rather than the $M_2$ term from the numerator of the 
intermediate neutrino propagator in the wave-function diagram. Hence, 
we do not account for this term here.

\subsection{Flavour Blind Interactions}
\label{sec:flblind}

The flavour-blind contribution to the lepton self-energy that
is mediated by gauge interactions can be expressed as
\begin{align}
\label{eq:sigmalblind}
{\rm i}{\Sigma\!\!\!\!/}_{\ell ab}^{{\rm bl}fg}
=g^2\int\frac{d^4 k^\prime}{(2\pi)^4}\frac{d^4 k^{\prime\prime}}{(2\pi)^4}
(2\pi)^4 \delta^4(k-k^\prime-k^{\prime\prime})
\gamma^\nu {\rm i}S_{\ell ab}^{fg}(k^\prime) \gamma^\mu
{\rm i}\Delta_{A\mu\nu}^{fg}(k^{\prime\prime})\,,
\end{align}
where ${\rm i}\Delta_{A\mu\nu}^{fg}$ is the gauge boson propagator on the CTP. The corresponding collision term is then 
%\begin{align}
%\label{coll:blind_full}
%{\cal C}_{\ell ab}^{\rm bl}(k)=&
%{\rm i}{\Sigma\!\!\!\!/}^{{\rm bl}>}_{\ell ac}(k) {\rm i}S_{\ell cb}^<(k)
%-{\rm i}{\Sigma\!\!\!\!/}^{{\rm bl}<}_{\ell ac}(k) {\rm i}S_{\ell cb}^>(k)
%= g^2\int\frac{d^4 k^\prime}{(2\pi)^4}\frac{d^4 k^{\prime\prime}}{(2\pi)^4}
%(2\pi)^4 \delta^4(k-k^\prime-k^{\prime\prime})
%\notag\\
%&\hspace{40mm}\times
%\Big[
%\gamma^\nu P_{\rm L} {\rm i}S_{\ell ac}^{>}(k^\prime) P_{\rm R} \gamma^\mu
%{\rm i}\Delta_{A\mu\nu}^{>}(k^{\prime\prime})P_{\rm L}{\rm i}S_{\ell cb}^<(k) P_{\rm R} 
%\notag\\
%&\hspace{40mm}\phantom{\times \Big[}
%- \gamma^\nu P_{\rm L} {\rm i}S_{\ell ac}^{<}(k^\prime) P_{\rm R} \gamma^\mu
%{\rm i}\Delta_{A\mu\nu}^{<}(k^{\prime\prime})P_{\rm L}{\rm i}S_{\ell cb}^>(k) P_{\rm R} \Big]\,.
%\end{align}
\begin{eqnarray}
\label{coll:blind_full}
{\cal C}_{\ell ab}^{\rm bl}(k ) &=\,&
{\rm i}{\Sigma\!\!\!\!/}^{{\rm bl}>}_{\ell ac}(k) {\rm i}S_{\ell cb}^<(k)
-{\rm i}{\Sigma\!\!\!\!/}^{{\rm bl}<}_{\ell ac}(k) {\rm i}S_{\ell cb}^>(k)
\\
&& \hspace*{-1.2cm} 
= \,g^2\int\frac{d^4 k^\prime}{(2\pi)^4}\frac{d^4 k^{\prime\prime}}{(2\pi)^4}
(2\pi)^4 \delta^4(k-k^\prime-k^{\prime\prime})
\,\Big[
\gamma^\nu {\rm i}S_{\ell ac}^{>}(k^\prime) \gamma^\mu
{\rm i}\Delta_{A\mu\nu}^{>}(k^{\prime\prime}){\rm i}S_{\ell cb}^<(k)  
\nonumber\\
&& \hspace*{-0.7cm}
- \,\gamma^\nu {\rm i}S_{\ell ac}^{<}(k^\prime) \gamma^\mu
{\rm i}\Delta_{A\mu\nu}^{<}(k^{\prime\prime}){\rm i}S_{\ell cb}^>(k) \Big]\,.
\nonumber
\end{eqnarray}

To check the consistency of the generalised chemical potential 
ansatz~(\ref{general:chem}), we now verify that the collision 
term (\ref{coll:blind_full}) vanishes provided the leptons and
anti-leptons have opposite chemical potentials, 
$\mu^-_{ab} = - \mu^+_{ab}$. To see this, we first note that 
$\mu^-_{ab} = - \mu^+_{ab}\equiv  - \mu_{ab}$ implies the generalised  
KMS relation $S_{\ell ab}^>(k) = - \left( e^{\beta k_0 - \beta \mu} 
\right)_{ac} S_{\ell cb}^<(k)$. Using this and the fact that the 
gauge bosons are in thermal equilibrium, which implies 
$\Delta_{A\mu\nu}^{>}(k) =  e^{\beta k_0} \Delta_{A\mu\nu}^{<}(k)$, 
in Eq.~(\ref{eq:sigmalblind}) yields the analogous relation 
${\Sigma\!\!\!\!/}_{\ell ab}^{{\rm bl}>} = - \left( e^{\beta k_0 - \beta \mu} 
\right)_{ac}{\Sigma\!\!\!\!/}_{\ell cb}^{{\rm bl}<}$ for the lepton 
self energy. This allows us to write 
\begin{equation}
{\cal C}_{\ell}^{\rm bl}(k ) =
\left[{\rm i}{\Sigma\!\!\!\!/}^{{\rm bl}<}_{\ell}(k),
e^{\beta k_0 - \beta \mu}\right]
\,{\rm i}S_{\ell}^<(k)
\end{equation}
Inserting Eq.~(\ref{eq:sigmalblind}) leaves the commutator 
$\left[{\rm i}{S}^{{\rm bl}<}_{\ell}(k),
e^{\beta k_0 - \beta \mu}\right]$ in flavour space, which is easily
seen to vanish upon use of the ansatz (\ref{Slessgreater:ansatz}) 
together with Eq.~(\ref{general:chem}) and $\mu^-_{ab} = - \mu^+_{ab}$.
The vanishing of the collision term under the conditions 
$\mu^-_{ab} = - \mu^+_{ab}$ means that the kinetic equilibrium 
distribution~(\ref{general:chem}) with opposite chemical potentials 
is indeed a stationary solution of the kinetic equation in the limit when
only the fast flavour-blind gauge interactions are present. 

Furthermore, when the gauge bosons are in equilibrium and the
equilibrium deviation of the leptons is small and parametrised as in 
Eq.~(\ref{delta:f}), we can approximate the collision term 
(\ref{coll:blind_full}) as
\begin{eqnarray}
\label{coll:blind}
{\cal C}_{\ell ab}^{\rm bl}(k)
&\approx&
g^2\int\frac{d^4 k^\prime}{(2\pi)^4}\frac{d^4 k^{\prime\prime}}{(2\pi)^4}
(2\pi)^4\delta^4(k-k^\prime-k^{\prime\prime})
\\\notag
&&\hspace*{-1.5cm} \times \,
\Big\{
\gamma^\nu {\rm i}\delta S_{\ell ab}(k^\prime) \gamma^\mu
\left[{\rm i}\Delta_{A\mu\nu}^>(k^{\prime\prime}) {\rm i}
S_{\ell bb}^<(k) 
-{\rm i}\Delta_{A\mu\nu}^<(k^{\prime\prime}) 
{\rm i}S_{\ell bb}^>(k) \right]
\\\notag
&&\hspace*{-1cm}
+\,\left[
\gamma^\nu {\rm i} S^>_{\ell aa}(k^\prime) \gamma^\mu
{\rm i}\Delta_{A\mu\nu}^>(k^{\prime\prime})
-
\gamma^\nu {\rm i} S^<_{\ell aa}(k^\prime) \gamma^\mu
{\rm i}\Delta_{A\mu\nu}^<(k^{\prime\prime})
\right]
{\rm i}\delta S_{\ell ab}(k) 
\Big\}
\,.
\end{eqnarray}
For this expression, we note that the terms in square brackets
are odd under a change of sign of the momenta, since 
to leading order in deviations from equilibrium, we may substitute the
equilibrium distributions and make use of the fact that
${\rm i}\Delta_{A\mu\nu}^{{\rm eq}>}(k)=
{\rm i}\Delta_{A\mu\nu}^{{\rm eq}<}(-k)$
and ${\rm i}S^{{\rm eq}>}_{\ell aa}(k)={\rm i}S^{{\rm eq}<}_{\ell aa}(-k)$.
Hence, after performing the $k^0$ integration of the collision term, 
the same sign contributions to the
equations (\ref{kin:eq:nu}) and  (\ref{kin:eq_nu}) 
for $\delta n_\ell^+$ and $\delta n_\ell^-$ occur due to a
cancellation of the relative sign in Eq.~(\ref{kin:eq:nu}). 

However, if we substituted tree-level propagators in Eq.~(\ref{coll:blind}),
this collision term would vanish, since
all the three particles involved are massless. It is therefore necessary to
account for thermal masses and for finite width effects, that relax
the zero-temperature on-shell conditions. When employing finite width
propagators, analytical simplifications of the collision integral due
to on-shell $\delta$-functions no longer apply. In the present work,
we therefore do not perform collision integrals that vanish for 
zero-temperature propagators explicitly. Rather,
we discuss their general form and give estimates, while relegating
more precise numerical evaluations to future studies.

Of particular interest within the collision term~(\ref{coll:blind}) are
contributions for which ${\rm sign} (k^{\prime\prime\,0})=
-{\rm sign} (k^{\prime\,0})={\rm sign} (k^0)$. 
These are allowed when we account for the finite width
in the spectral functions (for both, $\ell$ and the gauge fields $A$) and
they correspond to lepton anti-lepton 
pair creation and annihilation processes.
After performing the integrations and the Dirac trace, lepton- and antilepton
contributions are identical. Therefore, we may parametrise 
the flavour blind contribution to the collision term in
Eq.~(\ref{kin:eq:nu}) by
\begin{equation}
\pm\frac12\,{\rm tr}\int\limits_{0,-\infty}^{\infty,0}\frac{dk_0}{2\pi}
\int\frac{d^3 k}{(2\pi)^3}({\cal C}^{\rm bl}_{\ell ab}+{\cal C}^{\rm
  bl \,\dagger}_{\ell
  ab})
= - \Gamma^{\rm bl}\,(\delta n^+_{\ell ab}+\delta n^-_{\ell ab})\,,
\end{equation}
which leads to the corresponding term in Eq.~(\ref{kin:eq_nu}). Here 
$\delta n_{\ell ab}^\pm$ has been factored out by substituting 
Eqs.~(\ref{delta:f}) and ~(\ref{delta:S}).
Note that by use of Eqs.~(\ref{delta:f}) and~(\ref{delta:S}) this
defines $\Gamma^{\rm bl}$, which therefore may readily be evaluated
within a more detailed numerical study. For the purposes of the present work,
we estimate $\Gamma^{\rm bl}\sim g_2^4 T$, where
$g_2$ is the ${\rm SU}(2)_{\rm L}$ gauge coupling and where the
additional factor of $g_2^2$ compared to
the tree-level matrix element arises from the finite-width
effects~\cite{Garbrecht:2008cb}.

The fact that the flavour-blind collision terms for
$\delta n_\ell^+$ and $\delta n_\ell^-$ are of the same sign also
implies that in the absence of additional flavour-sensitive effects,
$\delta n_\ell^+-\delta n_\ell^-$ is conserved, as it is required
for the ansatz of generalised chemical potentials~(\ref{general:chem})
to be valid.

\subsection{Flavour Sensitive Interactions}

We now turn to the active lepton Yukawa couplings. These contribute
to the self-energy of the left-handed leptons as
\begin{align}
{\rm i}{\Sigma\!\!\!\!/}^{{\rm fl}fg}_{\ell ab}(k)=
h_{ac}^\dagger h_{db}\int\frac{d^4 k^\prime}
{(2\pi)^4}\frac{d^4 k^{\prime\prime}}{(2\pi)^4}
(2\pi)^4\delta^4(k-k^\prime-k^{\prime\prime})
\,{\rm i}S_{{\rm R}cd}^{fg}(k^\prime) 
{\rm i}\Delta_\phi^{fg}(k^{\prime\prime})\,.
\end{align}
To linear order in deviations from equilibrium, the collision term is
\begin{align}
{\cal C}_{\ell ab}^{\rm fl}(k)
=\,\,&{\rm i}{\Sigma\!\!\!\!/}^{{\rm fl}>}_{\ell ac}(k)
{\rm i}S_{\ell cb}^<(k)
-{\rm i}{\Sigma\!\!\!\!/}^{{\rm fl}<}_{\ell ac}(k)
{\rm i}S_{\ell cb}^>(k)
\approx
\int\frac{d^4 k^\prime}{(2\pi)^4}\frac{d^4 k^{\prime\prime}}{(2\pi)^4}
(2\pi)^4\delta^4(k-k^\prime-k^{\prime\prime})
\\\notag
&\times\Big\{
h_{ac}^\dagger h_{de}
{\rm i}\delta S_{{\rm R}cd}(k^\prime)
\left[
{\rm i}\Delta^>_\phi(k^{\prime\prime}) {\rm i}S^<_{\ell eb}(k)
-{\rm i}\Delta^<_\phi(k^{\prime\prime}) {\rm i}S^>_{\ell eb}(k)
\right]
\\\notag
&\hspace*{0.45cm}
+
h^\dagger_{ac} h_{de}\left[
{\rm i}S_{{\rm R}cd}^>(k^\prime){\rm i}\Delta_\phi^>(k^{\prime\prime})
-{\rm i}S_{{\rm R}cd}^<(k^\prime){\rm i}\Delta_\phi^<(k^{\prime\prime})
\right]
{\rm i}\delta S_{\ell eb}(k) 
\Big\}
\\\notag
=&
\int\frac{d^4 k^\prime}{(2\pi)^4}\frac{d^4 k^{\prime\prime}}{(2\pi)^4}
(2\pi)^4\delta^4(k-k^\prime-k^{\prime\prime})
\\\notag
&\times\Big\{
h_{ac}^\dagger h_{db}
{\rm i}\delta S_{{\rm R}cd}(k^\prime)
\left[
{\rm i}\Delta^>_\phi(k^{\prime\prime})
{\rm i}S^{{\rm eq}<}_{\ell}(k)
-{\rm i}\Delta^<_\phi(k^{\prime\prime})
{\rm i}S^{{\rm eq}>}_{\ell}(k)
\right]
\\\notag
&\hspace*{0.45cm}
+
h^\dagger_{ac} h_{ce}\left[
{\rm i}S_{{\rm R}}^{{\rm eq}>}(k^\prime){\rm i}
\Delta_\phi^>(k^{\prime\prime})
-{\rm i}S_{{\rm R}}^{{\rm eq}<}(k^\prime){\rm i}
\Delta_\phi^<(k^{\prime\prime})\right]
{\rm i}\delta S_{\ell eb}(k) 
\Big\}
\,.
\end{align}
Again, the leading thermal corrections to the propagators should be employed,
since at tree-level, this integral is vanishing for kinematic reasons.
We have to distinguish two relevant kinematic situations: First,
when ${\rm sign}(k^0)=-{\rm sign}(k^{\prime \,0})$,
the collision term corresponds to pair creation or annihilation of a left-
and a right-handed Standard Model lepton. Second, when
${\rm sign}(k^0)={\rm sign}(k^{\prime \,0})$ the left- and 
right-handed leptons scatter
from a Higgs boson. Again, both configurations are only possible due to the
finite width of the spectral functions of $\ell$, ${\rm R}$ and $\phi$.
We summarise both contributions  to Eq.~(\ref{kin:eq:nu}) by writing
\begin{align}
\label{Gamma:fl}
\Gamma^{\pm \rm fl}_{\ell ab}=&
\pm\frac12
{\rm tr}\int\limits_{0,-\infty}^{\infty,0}\frac{d k^0}{2\pi}\int\frac{d^3k}{(2\pi)^3}
\left(
{\cal C}_{\ell ab}^{\rm fl}(k)
+{\cal C}_{\ell ab}^{{\rm fl}\dagger}(k)
\right)
\\\notag
=&
\Gamma^{\rm an}
\left(
[h^\dagger h]_{ac} \delta n_{\ell cb}^{\pm}+
\delta n_{\ell ac}^{\pm \dagger}[h^\dagger h]_{cb}
+h_{ac}^\dagger \delta n_{{\rm R}cd}^{\mp} h_{db}
+h_{ad}^\dagger \delta n_{{\rm R}dc}^{\mp \dagger} h_{cb}
\right)
\\\notag
+&\Gamma^{\rm sc}
\left(
[h^\dagger h]_{ac} \delta n_{\ell cb}^{\pm}+
\delta n_{\ell ac}^{\pm \dagger}[h^\dagger h]_{cb}
-h_{ac}^\dagger \delta n_{{\rm R}cd}^{\pm} h_{db}
-h_{ad}^\dagger \delta n_{{\rm R}dc}^{\pm \dagger} h_{cb}
\right)\,.
\end{align}
For later use we note that for the right handed leptons, we have 
the corresponding flavour sensitive scattering rate
\begin{align}
\Gamma^{\pm \rm fl}_{{\rm R} ab}=&
\Gamma^{\rm an}
\left(
[h h^\dagger]_{ac} \delta n_{{\rm R} cb}^{\pm}+
\delta n_{{\rm R} ac}^{\pm \dagger}[h h^\dagger]_{cb}
+h_{ac} \delta n_{\ell cd}^{\mp} h_{db}^\dagger
+h_{ad} \delta n_{\ell dc}^{\mp \dagger} h_{cb}^\dagger
\right)
\\\notag
+&\Gamma^{\rm sc}
\left(
[h h^\dagger]_{ac} \delta n_{{\rm R} cb}^{\pm}+
\delta n_{{\rm R} ac}^{\pm \dagger}[h h^\dagger]_{cb}
-h_{ac} \delta n_{\ell cd}^{\pm} h_{db}^\dagger
-h_{ad} \delta n_{\ell dc}^{\pm \dagger} h_{cb}^\dagger
\right)\,.
\end{align}
We estimate the factors $\Gamma^{\rm an}$ and $\Gamma^{\rm sc}$ as
$\sim g_2^2 T$, which is again due to the finite width of $\ell$ and
$\phi$ at finite temperature. (The ${\rm U}(1)_Y$ contribution is
smaller because of the smaller gauge coupling and the
smaller number of gauge bosons).
We recall that within
these expressions, $\delta n_\ell^\pm$ and the second index of the
coupling $h$ transform under left-handed flavour rotations,
while $\delta n_{\rm R}^\pm$ and the first index of $h$ remain without
change.
By a unitary transformation of the right-handed flavour basis,
we may choose the matrix $hh^\dagger$ to be diagonal, which is what we assume here.

This concludes the derivation of the kinetic
equation~(\ref{kin:eq_nu}) for the number densities.
An analogous equation (without washout and source terms) holds for the
right-handed Standard Model leptons.

\subsection{Suppression of Flavour Oscillations}

The largest collision term within the kinetic equations~(\ref{kin:eq_nu})
is $\Gamma^{\rm bl}=O(g_2^4 T)$. Close to equilibrium,
it imposes the constraint
\begin{align}
\label{charge_constraint}
\delta n_{ab}^+=-\delta n_{ab}^-\,.
\end{align}
This is expected, since in the flavour-blind limit
where $h_{ab}\to 0$, this condition is manifestly invariant
with respect to flavour rotations and it reduces to the assumption that
the lepton charge density of leptons is the same as the lepton charge density
of anti-leptons. Therefore, this
condition is implicitly
employed in Ref.~\cite{Beneke:2010wd} as well
as in many other kinetic-theory approaches to various problems. Now,
due to the $\pm$ in the first term on the right hand side of the
kinetic equations~(\ref{kin:eq_nu}), a large $\Gamma^{\rm bl}$
effectively inhibits flavour oscillations, which would be present 
in the absence of collisions. To see this in more detail, consider
the toy system of differential equations
\begin{subequations}
\label{g_matrix_eq}
\begin{align}
\label{g_matrix_eq1}
\frac{d}{dt}\delta g^+(t)&=-{\rm i}\Delta\omega\delta g^+(t)-
\Gamma[\delta g^+(t)+\delta g^-(t)]\,,
\\
\label{g_matrix_eq2}
\frac{d}{dt}\delta g^-(t)&=+{\rm i}\Delta\omega \delta g^-(t)-
\Gamma[\delta g^-(t)+\delta g^+(t)]
\,.
\end{align}
\end{subequations}
The relevant parameters for flavoured leptogenesis can be estimated as
\begin{align}
\Gamma=\Gamma^{\rm bl}\sim g_2^4 T\,,\quad\Delta\omega\sim h_\tau^2 T 
\ll \Gamma\,,
\end{align}
where $h_\tau$ denotes the $\tau$-lepton Yukawa coupling.
Since $g_2^4 \gg h_\tau^2$,
the solutions are linear combinations of two eigenmodes with short
$\tau_{\rm s} = 1/(\Gamma+\sqrt{\Gamma^2-\Delta\omega^2}) \approx 1/(2 \Gamma)$ and long 
$\tau_{\rm l} = 1/(\Gamma-\sqrt{\Gamma^2-\Delta\omega^2}) \approx 2 \Gamma / \Delta\omega^2$
decay times, respectively. The corresponding eigenvectors are given by
\begin{align}
\delta g_{\rm s,l} = \delta g^+ +\frac{-{\rm i}
\Delta\omega \pm \sqrt{\Gamma^2-\Delta\omega^2}}{\Gamma} \delta g^-
\approx  \delta g^+ \pm \left(1 \mp {\rm i} 
\frac{\Delta\omega}{\Gamma} \right) \delta g^-\,,
\end{align}
with
\begin{align}
\delta g_{\rm s,l} = (\delta g_{\rm s,l})_0 \,{\rm e}^{-t/\tau_{\rm s,l}}\,.
\end{align}
The short mode $\delta g_{\rm s} \approx \delta g^+ +  \delta g^-$ is 
thus damped to zero very rapidly by pair annihilations, implying an 
effective constraint
\begin{align}
\label{eq:delg}
\delta g^+ \sim - \left(1 - {\rm i} \frac{\Delta\omega}{\Gamma}
\right) \delta g^-\,.
\end{align}
Note that the different sign of $\Delta \omega$ terms in 
Eq.~(\ref{g_matrix_eq}) is decisive, since it implies that 
the driving term for oscillations in 
\begin{equation}
\frac{d}{dt}\left(\delta g^+(t)-\delta g^-(t)\right)
=-{\rm i} \Delta\omega\left(\delta g^+(t)+\delta g^-(t)\right)
\end{equation}
is damped away, while in the case of same sign $\delta g^+-\delta g^-$,
could have freely oscillated. As explained in 
Section~\ref{section:massshells} the opposite sign of the 
$\Delta \omega$ term in Eq.~(\ref{kin:eq_nu}) is a consequence 
of $CP$ invariance at leading order.

Within the gradient expansion, the first order correction to 
Eq.~(\ref{charge_constraint})  therefore is of
order $\Delta\omega^{\rm eff}/\Gamma^{\rm bl}$.
Since the source terms for the off-diagonal correlations are already 
of first order in gradients, it is justified to use the zeroth order 
constraint Eq.~(\ref{charge_constraint}) within our approximations. 
The long-lived mode describes the damping of flavour coherence in the 
lepton charge density matrix due to 
flavour-blind interactions. It is much slower compared to the damping 
rate due to flavour sensitive interactions,
${\Delta\omega^{\rm eff}}^2/\Gamma^{\rm bl} \sim h_\tau^4 g_2^{-4} T
\ll \Gamma^{\rm fl}\sim g_2^2h_\tau^2 T$ since $h_\tau\ll g_2^3$. Therefore,
we may neglect the damping due to flavour-blind interactions, while
we keep the direct damping due to flavour sensitive processes.
While in the case of leptogenesis, we conclude that because of
$\Delta\omega\ll \Gamma$, flavour oscillations are overdamped
and effectively frozen,
we note that for $\Delta \omega > \Gamma$, there are damped
flavour oscillations. It is interesting to note that even though we 
assume flavour blind interactions, the off-diagonal  flavour coherence
functions are decaying. Such a behaviour, in particular in the 
oscillatory regime, has been observed numerically in 
Ref.~\cite{Cirigliano:2009yt}.

We emphasise that the conclusion that the oscillations induced by 
$\Delta\omega$
are overdamped for $\Gamma\gg \Delta\omega$ does not depend on the choice of
the flavour basis. To see this, we extend $g^\pm$ to a vector of 
functions and consider the system of matrix equations
\begin{subequations}
\label{g:toymod}
\begin{align}
\frac{d}{dt}\delta g^+(t)&=-{\rm i}[\omega,\delta g^+(t)]-\Gamma[\delta g^+(t)+\delta g^-(t)]\,,
\\
\frac{d}{dt}\delta g^-(t)&=+{\rm i}[\omega, \delta g^-(t)]-\Gamma[\delta g^-(t)+\delta g^+(t)]
\,.
\end{align}
\end{subequations}
Here, $\Gamma$ is proportional to the unit matrix and
$\omega=\omega^{\rm fl}+\omega^{\rm bl}$, where $\omega^{\rm bl}$ is
proportional to the unit matrix and $\omega^{\rm fl}_{ab}\ll\Gamma_{cc}$
for all $a,b,c$. It then follows that
$[\omega,\delta g^\pm(t)]=[\omega^{\rm fl},\delta g^\pm(t)]$.
By taking the sum of Eqs.~(\ref{g:toymod}), we again conclude that
$\delta g^+(t)+\delta g^-(t)\sim {\rm e}^{-2\Gamma t}$. Consequently,
the difference of Eqs.~(\ref{g:toymod}) yields
\begin{align}
\label{deltag:fixedbasis}
\frac{d}{dt}\left[\delta g^+(t)-\delta g^-(t)\right]
=0+
\left[\delta g^+(t)-\delta g^-(t)\right]\times{\cal O}\left(\frac{\omega^2_{ab}}{\Gamma_{cc}}\right)\,,
\end{align}
where the right hand side is estimated as the eigenvalues of a matrix 
with large diagonal and small off-diagonal elements.
Alternatively, this can be seen by substituting in the right hand side
of the difference of Eqs.~(\ref{g:toymod})
\begin{align}
\left[\omega,\delta g^+(t)+\delta g^-(t)\right]
=
\left(
\delta g^+(t)-\delta g^-(t)
\right)
\times {\cal O}\left(\omega^2/\Gamma\right)
\,,
\end{align}
where an estimate according to Eq.~(\ref{eq:delg}) is made.
This confirms the suppression of the
effect of $\varsigma^{\rm fl}$ by $\Gamma^{\rm bl}$ in a general flavour-basis.

We can also generalise this discussion to the case of a time-dependent mass
basis. In order to model this situation, consider the system 
\begin{subequations}
\begin{align}
\frac{d}{dt}\delta g_{ab}^+(t)
&=-{\rm i}\Delta\omega_{ab}\delta g_{ab}^+(t)
+\Xi_{ac} \delta g_{cb}^+ - \delta g_{ac}^+ \Xi_{cb}
-\Gamma[\delta g_{ab}^+(t)+\delta g_{ab}^-(t)]\,,
\\
\frac{d}{dt}\delta g_{ab}^-(t)
&=+{\rm i}\Delta\omega_{ab} \delta g_{ab}^-(t)
+\Xi_{ac} \delta g_{cb}^- - \delta g_{ac}^- \Xi_{cb}
-\Gamma[\delta g_{ab}^-(t)+\delta g_{ab}^+(t)]
\,.
\end{align}
\end{subequations}
In the limit $\Gamma \gg \Delta \omega_{ab}$, we may find an approximate solution by
imposing $\delta g^+=-\delta g^-$. This leads to
\begin{align}
\frac{d}{dt}\left(\delta g_{ab}^+(t)-\delta g_{ab}^-(t)\right)
=\left[\delta g_{ab}^+(t)-\delta g_{ab}^-(t),\Xi\right]\,,
\end{align}
which is solved by the unitary evolution
\begin{align}
\delta g_{ab}^+(t)-\delta g_{ab}^-(t)=
\left(
T{\rm e}^{-\Xi t}
\right)
\left(
\delta g_{ab}^+(t=0)-\delta g_{ab}^-(t=0)
\right)
\left(
\bar T
{\rm e}^{\Xi t}
\right)\,,
\end{align}
where $T$ implies the time-ordered exponential.
Therefore, the freezing of flavour oscillations also persists when we
account for the time dependence of the mass basis. From above equation, we recover
Eq.~(\ref{deltag:fixedbasis}) by undoing the flavour rotation, that is
by left multiplication by $U$ and right multiplication by $U^\dagger$.
% We thus observe that when neglect the flavour-dependent term
% $\varsigma^{\rm fl}$, the resulting kinetic equations are manifestly
% covariant under flavour rotations.

\subsection{Kinetic Equations for Left and Right Handed Number Densities}

We now define the charge number density matrix as
\begin{align}
\label{charge:den:matrix}
q_{\ell ab}=\delta n^+_{\ell ab}-\delta n^-_{\ell ab}\,.
\end{align}
Imposing that fast
(compared to the interactions accounted for in $W_{ab}$ and $S_{ab}$)
pair creating and annihilating
interactions enforce the constraint
\begin{align}
\delta n_{\ell ab}^+=-\delta n_{\ell ab}^-\,,
\end{align}
we can take the linear combinations from Eq.~(\ref{kin:eq_nu}) that solve
for the charge density matrix~(\ref{charge:den:matrix}). 
For the flavour-sensitive interactions, define
\begin{align}
\Gamma^{\rm fl}_{\ell ab}=&
\Gamma^{\rm an}
\left(
[h^\dagger h]_{ac} q_{\ell cb}+
q_{\ell ac}^{\dagger} [h^\dagger h]_{cb} 
-h^\dagger_{ac} q_{{\rm R}cd}h_{db}
-h^\dagger_{ad} q_{{\rm R}dc}^{\dagger} h_{cb}
\right)
\\\notag
+&\Gamma^{\rm sc}
\left(
[h^\dagger h]_{ac} q_{{\ell} cb}
+  q_{{\ell} ac}^{\dagger}[h^\dagger h]_{cb}
-h^\dagger_{ac} q_{{\rm R}cd}h_{db}
-h^\dagger_{ad} q_{{\rm R}dc}^{\dagger} h_{cb}
\right)\,. 
\end{align}
Using this and the results of the previous sections, we obtain the kinetic equations~(\ref{kin:eq}) which we repeat here for completeness:
\begin{align}
\label{kin:eq2}
%\notag
\frac{\partial q_{\ell ab}}{\partial \eta}
=
%-{\rm i}\Delta\omega^{\rm eff}_{\ell ab} q_{\ell ab}
\sum\limits_{c}\left[
q_{\ell ac}\Xi_{cb}
-\Xi_{ac}q_{\ell cb}
-W_{ac} q_{\ell cb}
-q_{\ell ac}W_{cb}
\right]
+ 2 S_{ab}
-\Gamma_{\ell ab}^{\rm fl}
\,.
\end{align}
Note that similar to the toy system of equations the flavour-blind 
term drops out in the equation for $q_{\ell ab}$, while it is consistent 
to neglect the $\Delta\omega_{ab}^{\rm eff}$ term, which would 
multiply $\delta n^+_{\ell ab}+\delta n^-_{\ell ab}$ in this equation,
which is strongly damped. This holds in an arbitrary, 
time-independent basis in flavour space, where the $\Xi$ terms 
are absent. In a time-dependent basis such as the basis where 
$\varsigma_{ab}$ is diagonal, the $\Xi$ terms are introduced to 
account for the time-dependent basis rotation.

For the right-handed leptons, there is the analogous equation
\begin{align}
\label{kin:eq:R}
\frac{\partial q_{{\rm R}ab}}{\partial \eta}=-\Gamma^{\rm fl}_{{\rm R}ab}
\end{align}
with
\begin{align}
\Gamma^{\rm fl}_{{\rm R} ab}=&
\Gamma^{\rm an}
\left(
[hh^\dagger]_{ac} q_{{\rm R} cb}+
q_{{\rm R} ac}^{\dagger} [hh^\dagger]_{cb}
-h_{ac} q_{\ell cd}h^\dagger_{db}
-h_{ad} q_{\ell dc}^{\dagger} h^\dagger_{cb}
\right)
\\\notag
+&\Gamma^{\rm sc}
\left(
[hh^\dagger]_{ac} q_{{\rm R} cb}
+  q_{{\rm R} ac}^{\dagger}[hh^\dagger]_{cb}
-h_{ac} q_{\ell cd}h^\dagger_{db}
-h_{ad} q_{\ell dc}^{\dagger} h^\dagger_{cb}
\right)\,.
\end{align}

Similar results have been obtained earlier within an approach that
makes use of the density matrix in an occupation number basis.
In its details,
the equation for the difference between lepton and anti-lepton densities
in Ref.~\cite{Abada:2006fw} exhibits however differences to our kinetic
equation~(\ref{kin:eq2}). It is not clear whether the lepton charge
densities in Ref.~\cite{Abada:2006fw} should correspond to our 
$q_{\ell ab}$ (which is the difference of the lepton density and the transpose
of the anti-lepton density) or to the difference of the lepton density and the 
anti-lepton density. In the former case, the flavour oscillations frequencies
in Ref.~\cite{Abada:2006fw} should have opposite signs for particle and 
transposed antiparticle modes, if they were to agree with our
result obtained within the CTP formalism. This is apparently not the situation
within the equation for the lepton charge density in Ref.~\cite{Abada:2006fw}.
In the latter case, as it follows from Eq.~(\ref{S:CP}) and the
discussion at the end of Section~\ref{section:massshells},
within the CTP formalism the washout and source matrices for the lepton 
and anti-lepton densities are transposed (or complex conjugated, as these 
matrices are Hermitian) with respect to each other, which is apparently 
not the case in Ref.~\cite{Abada:2006fw}. Furthermore, the same 
conclusions on damping of coherence would result from our equations 
if the charge density matrix were defined as
$\delta n^+_{\ell ab}-\delta n^-_{\ell ba}$.
Hence, with either 
interpretation, there is a difference between the occupation number 
formalism result that is derived in Ref.~\cite{Abada:2006fw} and
the kinetic equation~(\ref{kin:eq}) derived within the CTP formalism.
The phenomenological consequence of this can be seen when comparing the
present work with Ref.~\cite{De Simone:2006dd}, where the kinetic 
equations from Ref.~\cite{Abada:2006fw} are solved numerically. 
While in the present work,
we conclude that flavour oscillations effectively freeze out due to fast pair
creation and annihilation processes, the results in
Refs.~\cite{Abada:2006fw,De Simone:2006dd} imply
that the flavour oscillations are important and in particular faster than the
flavour-sensitive damping processes.

\section{Solutions to the Flavoured Kinetic Equations}
\label{section:numerics}

We are considering a scenario with two lepton flavours and
assume that there is one dominant Standard Model Yukawa coupling
$h_\tau$.
In the basis where the lepton Yukawa coupling matrix is 
diagonal, the matrix $h$ is therefore simply
\begin{align}
h=
\left(
\begin{array}{cc}
h_\tau & 0\\
0 & 0
\end{array}
\right)\,.
\end{align}
Provided the $\mu$ and $e$ Yukawa-couplings are negligible 
$ h^2_{\mu, e} T/a(\eta)\ll H$, the realistic case with three lepton 
flavours can be reduced to the present case by separating out a linear
combination of lepton flavours, for which no asymmetry is
produced. This corresponds to an unflavoured approximation for
the $e$ and $\mu$ flavours, {\it cf.} the discussion of the
unflavoured limit below.
We note that Eq.~(\ref{kin:eq}) is manifestly invariant under
flavour rotations induced by $U$, while Eqs.~(\ref{kin:eq:f})
and~(\ref{kin:eq:nu}) are not, because the term that 
describes flavour oscillations is given in the diagonal basis.
In our approximation, we can drop this term,
because we have shown in Section~\ref{sec:kinetic}
that due to the constraints from kinetic equilibrium,
the time scale for flavour oscillations is suppressed when compared
to the time scale of decoherence from flavour-sensitive scatterings.
We use this freedom  of choice of a lepton flavour basis
and perform the discussion in this section within the time-independent
basis of charged lepton flavours, which is more transparent
than the time-dependent basis of the leptonic quasi-particles, 
that is determined by the diagonalisation of $\varsigma^{\rm fl}$.
Likewise, we
present all numerical results in the basis of charged
lepton flavours.\footnote{Yet, we have used the requirement that computations in
both bases must yield the same results as a consistency check on the
numerical results.} 

In the charged lepton basis, the
flavour-sensitive collision terms read
\begin{subequations}
\begin{align}
\Gamma_\ell^{\rm fl}
=\left(\Gamma^{\rm an}+\Gamma^{\rm sc}\right)h_\tau^2
\left[
\left(
\begin{array}{cc}
1 & 0\\
0 & 0
\end{array}
\right)q_\ell
+q_\ell
\left(
\begin{array}{cc}
1 & 0\\
0 & 0
\end{array}
\right)
-
2\left(
\begin{array}{cc}
q_{{\rm R}11} & 0\\
0 & 0
\end{array}
\right)
\right]\,,\\
\Gamma_{\rm R}^{\rm fl}
=\left(\Gamma^{\rm an}+\Gamma^{\rm sc}\right)h_\tau^2
\left[
\left(
\begin{array}{cc}
1 & 0\\
0 & 0
\end{array}
\right)q_{\rm R}
+q_{\rm R}
\left(
\begin{array}{cc}
1 & 0\\
0 & 0
\end{array}
\right)
-
2\left(
\begin{array}{cc}
q_{\ell 11} & 0\\
0 & 0
\end{array}
\right)
\right]\,.
\end{align}
\end{subequations}
In the fully flavoured limit, which we define by the requirement
$(\Gamma^{\rm an}+\Gamma^{\rm fl})h_\tau^2\gg H$ (note that 
for the present purposes, ``fully flavoured''  refers to the
situation where $h_\mu$ and $h_e$ are still assumed to be
out-of-equilibrium), we see that within this
setup, the flavour sensitive collision terms enforce
\begin{align}
q_{\ell 11}-q_{{\rm R} 11}=0\,,\qquad
q_{\ell 12}=q_{\ell 21}=q_{{\rm R} 12}=q_{{\rm R} 21}=0\,.
\end{align}
This agrees with the expectation that when Standard Model 
Yukawa couplings are in equilibrium, the lepton asymmetries are
projected onto the charged lepton basis.

The processes $\ell + \bar {\rm R}\leftrightarrow \phi^*$ (annihilation) and
$\ell +\phi \leftrightarrow {\rm R}$ (scattering)
are kinematically forbidden when all the three particles involved are massless.
At finite temperature, this holds no longer true due to effects that at leading order
can be either thought of as thermal masses and finite widths or as radiation of
gauge bosons. The latter point of view is taken in Ref.~\cite{Joyce:1994zn} to calculate 
$\Gamma^{\rm sc}$. However, important $t$-channel diagrams are not included there and
a calculation of $\Gamma^{\rm an}$ is not provided. A systematic calculation
of these rates may be performed along the lines of Ref.~\cite{Arnold:2000dr}, which is
however beyond the scope of the current work. Motivated by the partial
result of Ref.~\cite{Joyce:1994zn}, we take here for numerical definiteness the estimate
\begin{align}
\Gamma^{\rm an}+\Gamma^{\rm sc}\approx 0.7 \alpha_{\rm W} T/a(\eta)
=1.75\times 10^{-2} \,T/a(\eta)\,,
\end{align}
which should be accurate up to a factor of order unity.
Besides, we take here $\alpha_{\rm W}=1/40$ as the weak coupling constant
at the scale of about $10^{12}\,{\rm GeV}$. The precise value depends
on the particular extension of the Standard Model.

To obtain numerical solutions, we first solve the kinetic equations
for the distribution of the right-handed neutrinos $N_1$.
They are given in Ref.~\cite{Beneke:2010wd}, and the generalisation from the
single flavour to the two-flavour case follows by the straightforward
replacement $|Y_1|^2\to \sum_a|Y_{1a}|^2$.
We employ this distribution
to calculate the washout and the source terms within
Eq.~(\ref{kin:eq}). To be specific, we choose
thermal initial conditions for $N_1$.
For the singlet neutrino masses, we choose $M_1=10^{12}\,{\rm GeV}$
and $M_2=10^{14}\,{\rm GeV}$. For the Yukawa couplings of the right handed
neutrinos, we consider two scenarios
\begin{align}
Y=
\left(
\begin{array}{cc}
1.4\times 10^{-2}    &    \;1 \times 10^{-2}\\
{\rm i} \times 10^{-1} &  \;10^{-1}
\end{array}
\right)
\,,\qquad \textnormal{Scenario~(A)}\,,
\\
Y=
\left(
\begin{array}{cc}
1.4\times 10^{-2}    &    \;3 \times 10^{-3}\\
{\rm i} \times 10^{-1} &  \;10^{-1}
\end{array}
\right)
\,,\qquad \textnormal{Scenario~(B)}\,.
\notag
\end{align}
We vary the Yukawa coupling $h_{\tau}$, since this
will directly exhibit the dependence of the results on the flavour effects,
while of course, for a phenomenological study, it would be more pertinent
to vary the unknown parameters $Y$ and $M_{1,2}$.

\begin{figure}
\begin{center}
\begin{tabular}{c}
\epsfig{file=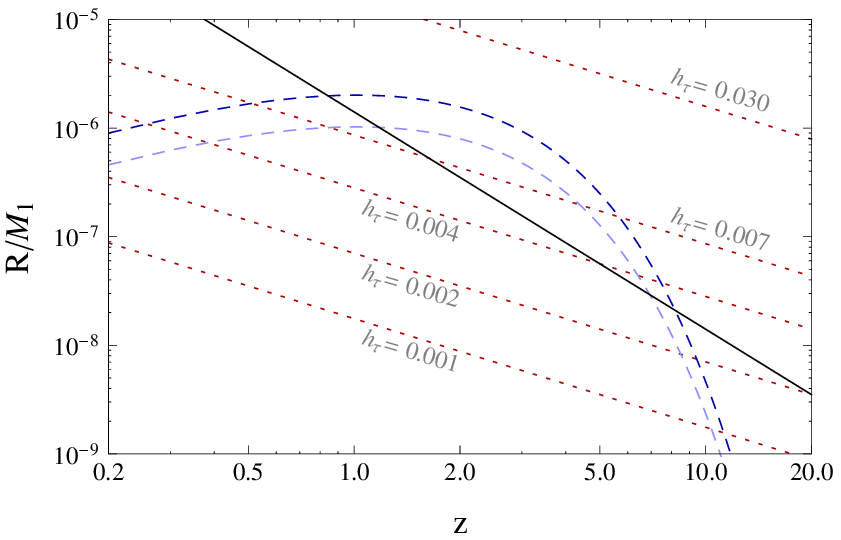,width=7cm}
\\
(A)
\end{tabular}
\begin{tabular}{c}
\epsfig{file=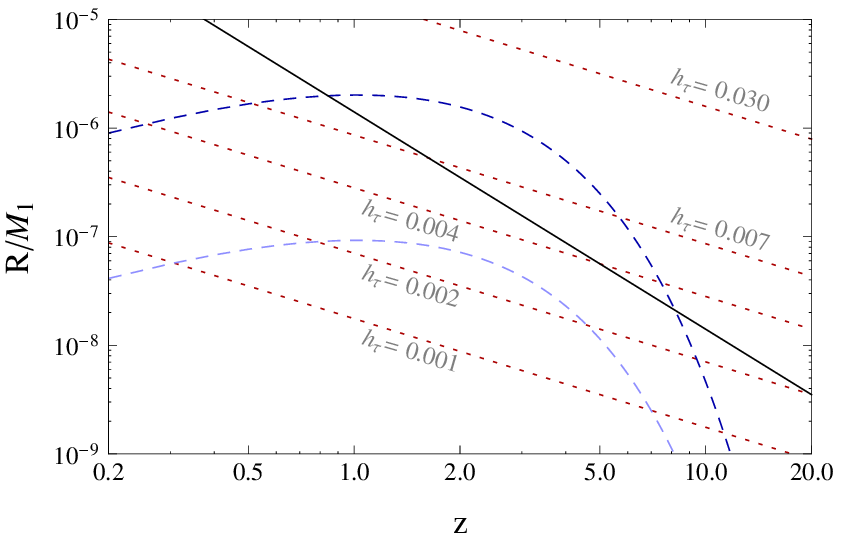,width=7cm}
\\
(B)
\end{tabular}
\end{center}
\caption{
\label{fig:rates}
Comparison of the relevant rates
$R=\Gamma^a_{\rm ID},\,H,\,h_\tau^2(\Gamma^{\rm an}+\Gamma^{\rm sc})$
for Scenarios~(A) and ~(B). The colour key is $H$ (solid, black),
$\Gamma^1_{\rm ID}$ (dashed, dark blue), $\Gamma^2_{\rm ID}$ (dashed, light blue), $h_\tau^2(\Gamma^{\rm an}+\Gamma^{\rm sc})$
with
$h_\tau=3\times 10^{-2},\;7\times 10^{-3},\;4\times 10^{-3},\;
2\times 10^{-3},\;10^{-3}$ (from top right to bottom left, dotted, red).
}
\end{figure}

% We expect a fully flavoured description to be applicable when
% $\Gamma^{\rm fl}\stackrel{>}{{}_\sim}H$ and
% $\Gamma^{\rm fl}\stackrel{>}{{}_\sim}\Gamma^{\rm ID}$, where we define the inverse
% decay rate as $\Gamma_{ID}=2(W_{11}+W_{22})$ with equilibrium distributions
% for the singlet neutrinos substituted within the washout terms. In Figure~\ref{fig:rates},
% we illustrate these rates for the case of Scenario~(A). This is a scenario with strong washout in both flavors, where the dominant contributions to the lepton asymmetry
% are generated between $z\approx 1$ and the point when the lepton
% asymmetry freezes out, $\Gamma_{ID}\approx H$. We see that during strong washout,
% the fully flavoured description is justified when $\Gamma^{\rm fl}\stackrel{>}{{}_\sim}\Gamma_{\rm ID}$ at all times during leptogenesis, whereas
% the unflavoured treatment is valid provided
% $\Gamma^{\rm fl}\stackrel{<}{{}_\sim}H \approx \Gamma_{\rm ID}$ at
% freeze-out~\cite{Blanchet:2006ch}. When neither of these conditions hold,
% we are in an intermediate regime that can neither be described through the
% unflavoured nor the fully flavoured approach. 

\begin{figure}
\begin{center}
\epsfig{file=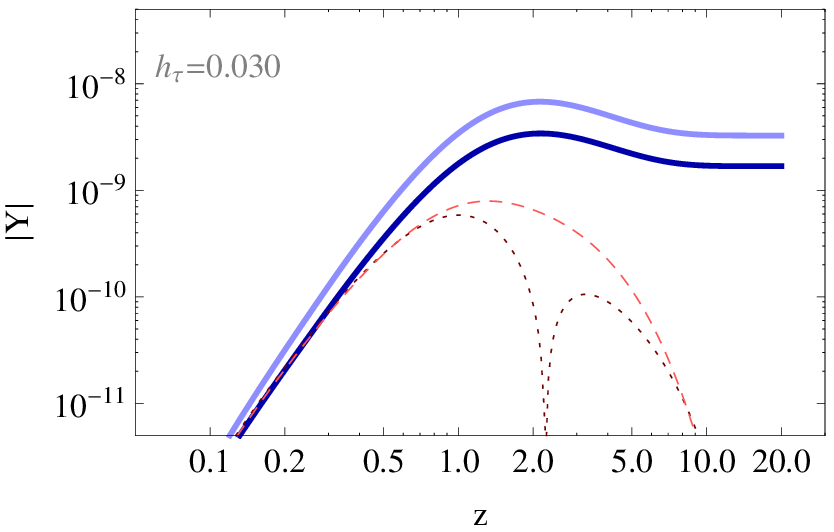,width=7.5cm}
\hskip0.5cm
\epsfig{file=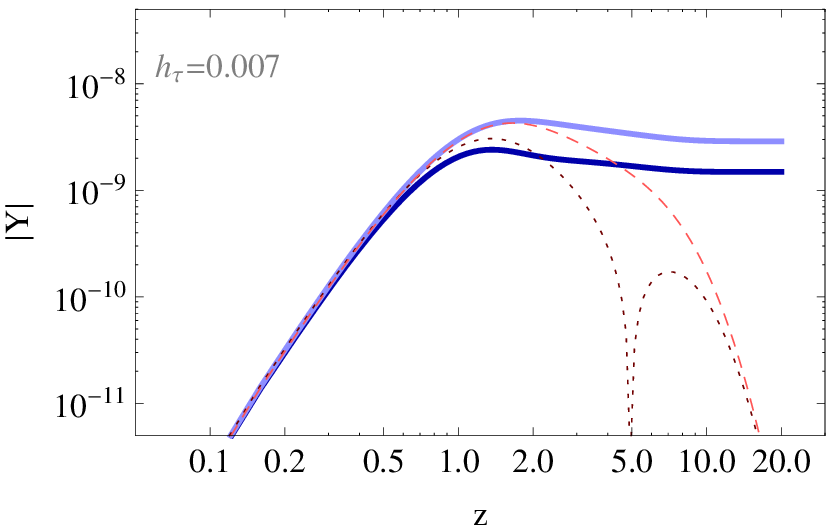,width=7.5cm}
%\hskip0.cm
\end{center}
\begin{center}
\epsfig{file=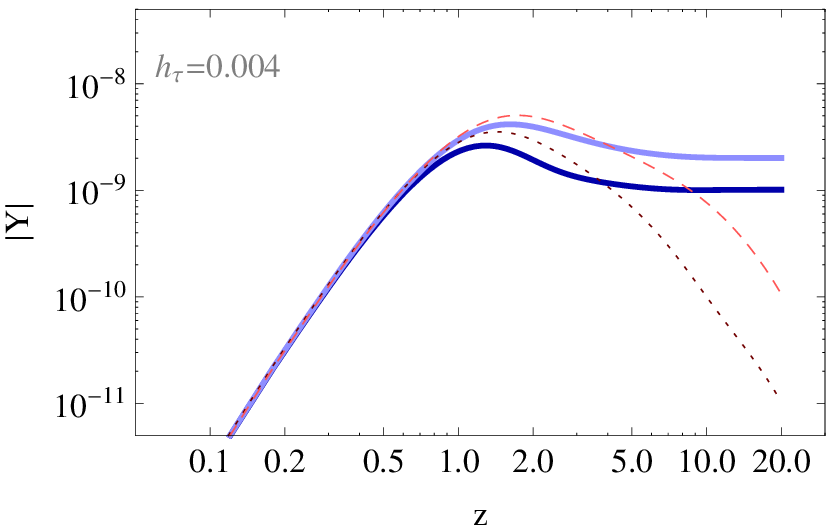,width=7.5cm}
\hskip0.5cm
\epsfig{file=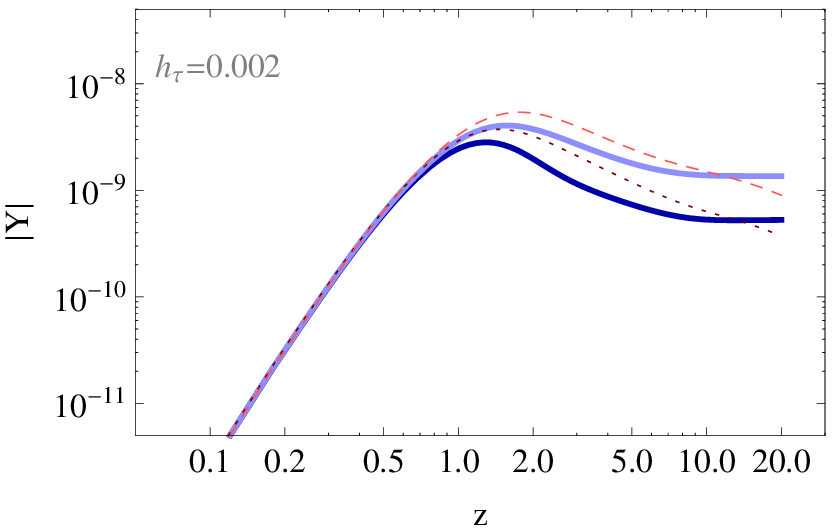,width=7.5cm}
\end{center}
\begin{center}
\epsfig{file=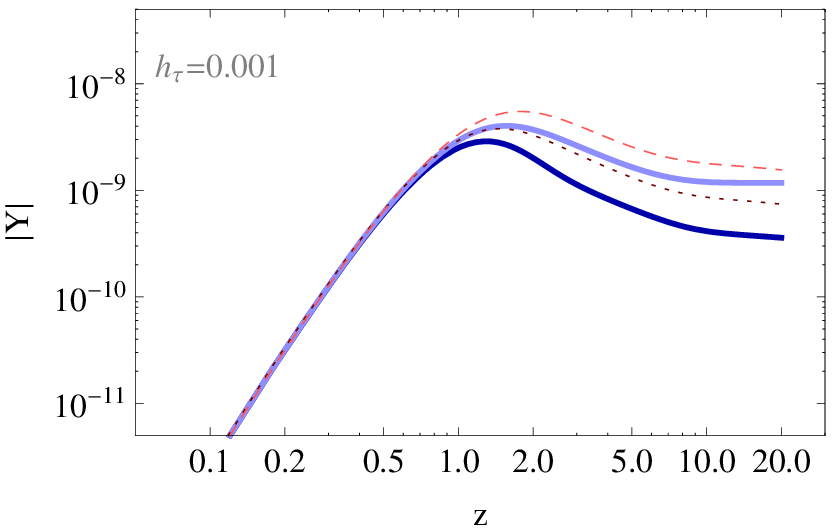,width=7.5cm}
\hskip0.5cm
\epsfig{file=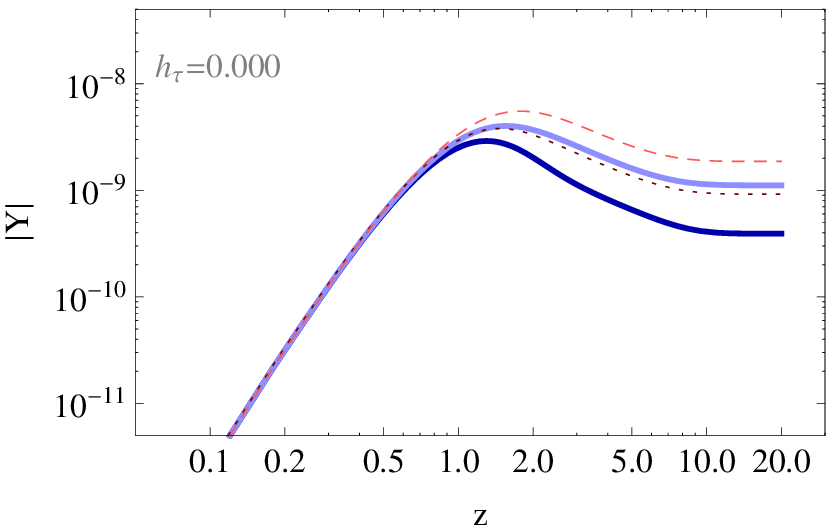,width=7.5cm}
\end{center}

\caption{
\label{fig:strongstrong}
Results for Scenario~(A) with 
$h_\tau=3\times 10^{-2},\;7\times 10^{-3},\;
4\times 10^{-3},\;2\times 10^{-3}\;,10^{-3},\;0$, from top left to bottom right. The key is
$Y_{\ell 11}$ (dark blue, solid), $Y_{\ell 22}$ (light blue, solid), ${\rm Re}[Y_{\ell 12}]$ (dark red, dotted),
 ${\rm Im}[Y_{\ell 12}]$ (light red, dashed). The densities are evaluated in the flavour
eigenbasis, which means that the larger the flavour effects are (the larger $h_\tau$ is),
the smaller are the off-diagonal densities.
}
\end{figure}

\begin{figure}

\begin{center}
\epsfig{file=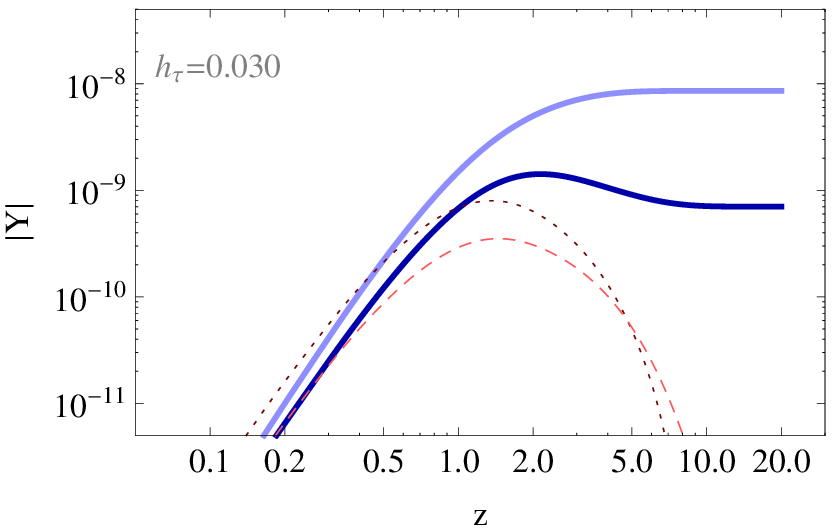,width=7.5cm}
\hskip0.5cm
\epsfig{file=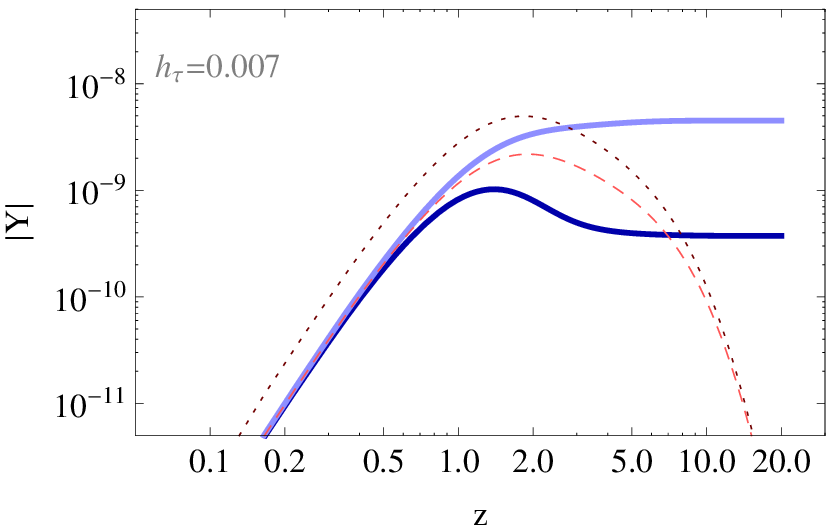,width=7.5cm}
%\hskip0.cm
\end{center}
\begin{center}
\epsfig{file=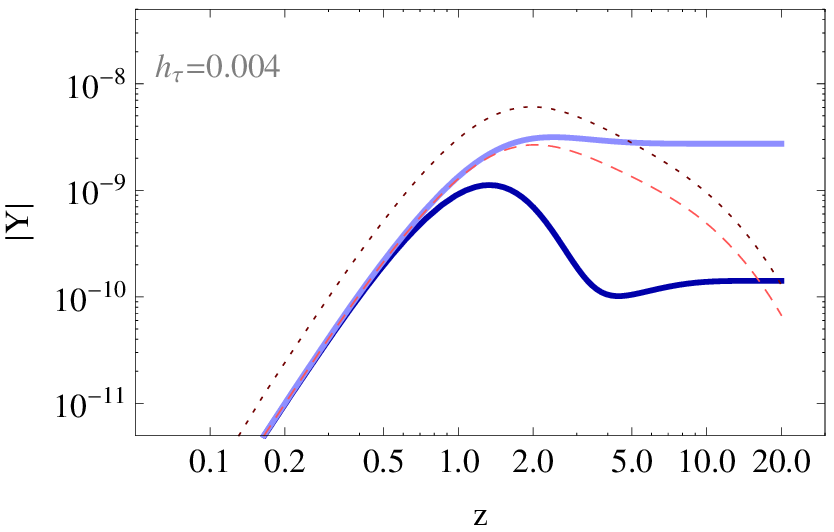,width=7.5cm}
\hskip0.5cm
\epsfig{file=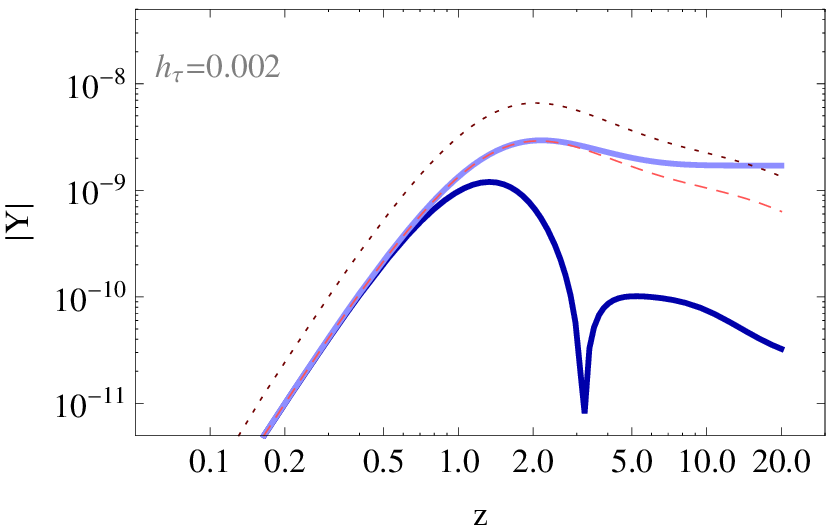,width=7.5cm}
\end{center}
\begin{center}
\epsfig{file=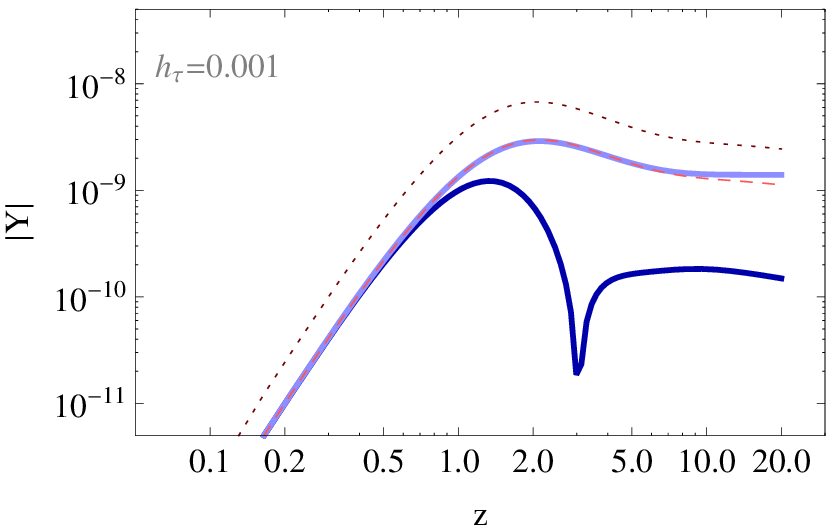,width=7.5cm}
\hskip0.5cm
\epsfig{file=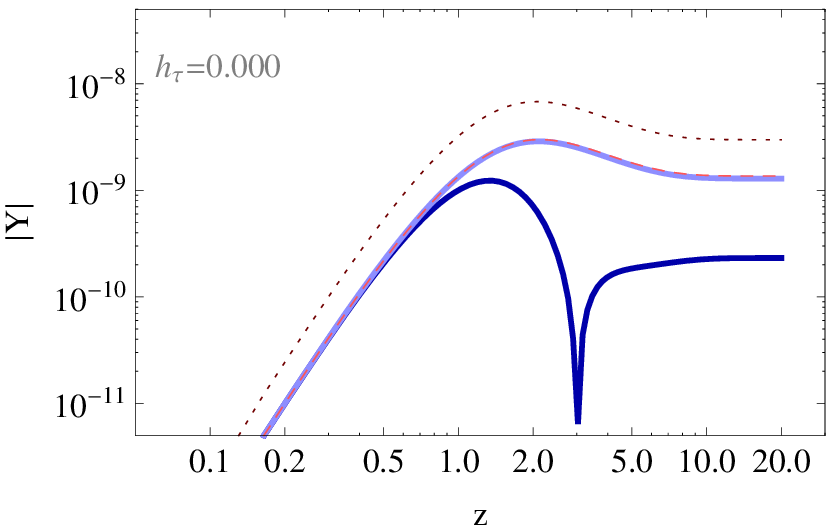,width=7.5cm}
\end{center}

\caption{
\label{fig:weakstrong}
Results for Scenario~(B) with
$h_\tau=3\times 10^{-2},\;7\times 10^{-3},\;
4\times 10^{-3},\;2\times 10^{-3}\;,10^{-3},\;0$ from top left to
bottom right. The key is
$Y_{\ell 11}$ (dark blue, solid), $Y_{\ell 22}$ (light blue, solid), ${\rm Re}[Y_{\ell 12}]$ (dark red, dotted),
 ${\rm Im}[Y_{\ell 12}]$ (light red, dashed).
}
\end{figure}

In Figure~\ref{fig:rates} we show the interaction rates $\Gamma^{\rm fl} = h_\tau^2 (\Gamma^{\rm an} + \Gamma^{\rm sc}) $ for different values of $h_\tau$ and compare them to the expansion rate of the universe $H$ and to the inverse decay rate for the individual flavours $a$,
$\Gamma^a_{\rm ID} = 2W_{aa}$ as a function of the ratio of $M_1$ to the physical temperature,
$z=a(\eta) M_1/T$.
Scenario~(A) exhibits moderate to strong washout in both flavours, where the dominant
contributions to the lepton asymmetry
are generated between $z\approx 3$ and the point when the lepton
asymmetry freezes out, $\Gamma_{ID}\approx H$.
We expect flavour effects to be negligible when $\Gamma^{\rm fl}\stackrel{<}{{}_\sim}H \approx \Gamma^a_{\rm ID}$ during
these times before freeze out~\cite{Blanchet:2006ch}, i.e. for $h_\tau$ significantly smaller
than $4\times10^{-3}$ by inspection of Figure~\ref{fig:rates}.
On the other hand a fully flavoured description should be applicable when
$\Gamma^{\rm fl}\stackrel{>}{{}_\sim}\Gamma_{\rm ID}^a$
during the times when the quantitatively relevant contributions to the
lepton asymmetry are produced, i.e. 
for  $h_\tau$  significantly larger than $7\times 10^{-3}$.
The numerical solutions to the kinetic equations for Scenario~(A)
are displayed in Figure~\ref{fig:strongstrong}. Shown are the absolute
values of the entropy normalised asymmetries
\begin{align}
    Y_{\ell ab} & = 2 g_w \frac{q_{\ell ab}}{\frac{2\pi^2}{45}g_\star T^3}\,,
\end{align}
where we use $g_\star=106.75$.
Since $Y_\ell$ is Hermitian, we plot the real and imaginary parts of $Y_{\ell 12}$. 
The results confirm our expectations about the validity of the fully 
flavoured and the unflavoured descriptions of leptogenesis. For 
$h_\tau \stackrel{<}{{}_\sim} 2\times 10^{-3}$ the total lepton 
asymmetry ${\rm tr} [Y_\ell] = Y_{\ell 11} +Y_{\ell 22}$ is almost 
independent of $h_\tau$, and the off diagonal densities decay away after
freeze out ($z\approx10$) only. (Only the initial stages of this decay
are visible in the plots for $h_\tau=2\times 10^{-3},\,10^{-2}$ in 
Figure~\ref{fig:strongstrong} due to the cut off at $z=20$. 
For $h_\tau=0$, the off-diagonal densities do not decay.)
On the other hand in the fully flavoured regime, for 
$h_\tau \stackrel{>}{{}_\sim} 3\times 10^{-2}$, the off diagonal 
densities are strongly suppressed before freeze-out.
This confirms that neglecting the off-diagonal densities, an
approximation that is commonly used in the fully flavoured regime, 
is indeed justified in this regime. 
In the intermediate regime, where neither of 
these approximations is valid, the correct lepton asymmetry is 
obtained by solving the full kinetic equation (\ref{kin:eq}). 

% Comparison with
% the rates in Figure~\ref{fig:rates} indeed confirms our expectations about
% the validity of the flavoured and the unflavoured descriptions of leptogenesis.

Scenario~(B) is a situation with strong washout due to
$Y_{11}$ and weak washout for $Y_{12}$. The numerical solutions
are displayed in Figure~\ref{fig:weakstrong}.
Since $\Gamma^2_{\rm ID}$ is now significantly smaller, it takes
also smaller values of $h_\tau$ before the unflavoured description
may be expected to be valid, {\it cf.} Figure~\ref{fig:rates}.
In the fully flavoured regime, we observe that
one of the lepton flavours suffers from strong washout while the other one
is only weakly washed out. On the other hand, in the basis where 
the source term is diagonal, the flavours apparently
mix in such a way that both lepton flavours are strongly washed out when flavour effects are turned off.
This importance of flavour effects for washout
is well known~\cite{Pilaftsis:2005rv,Abada:2006fw,Nardi:2006fx},
and it can be easily understood when
we recall how the fully flavoured and the unflavoured regimes are described.

First, in the fully flavoured case, densities that are off-diagonal in
the flavour
basis undergo fast damping through flavour-sensitive interactions. As a result,
there are two washout rates that are proportional to $|Y_{11}^2|$ and
$|Y_{12}^2|$, respectively. Second, in the unflavoured regime, it is 
convenient to bring $Y$ to a triangular form through
\begin{align}
Y^\Delta_{ia}=Y_{ib} V_{ba}\,,
\end{align}
where
\begin{align}
V=\frac{1}{\sqrt{|Y_{21}|^2+|Y_{22}|^2}}\left(
\begin{array}{cc}
Y_{22} & Y_{21}^*\\
-Y_{21} & Y_{22}^*
\end{array}
\right)\,.
\end{align}
In the triangular basis, a lepton asymmetry is only produced for the
linear combination
\begin{align}
\frac{1}{\sqrt{|Y_{21}|^2+|Y_{22}|^2}}\left(
Y_{21}\ell_1 + Y_{22} \ell_2
\right)\,.
\end{align}
The  washout rate for this linear combination is proportional to
\begin{align}
|Y_{12}^\Delta|^2=\frac{1}{|Y_{21}|^2+|Y_{22}|^2}
\left(
|Y_{11}|^2|Y_{21}|^2+|Y_{12}|^2|Y_{22}|^2+
2{\rm Re}\left[Y_{11}Y_{21}^*Y_{12}^* Y_{22}\right]
\right)\,.
\end{align}
The change of the effective flavour basis in the transition from the 
unflavoured to the flavoured regime therefore explains the apparent 
change in the washout rates for the individual flavours
for Scenario~(B), that are visible in Figure~\ref{fig:weakstrong}.

To obtain an estimate of the extent of the intermediate regime in
terms of the parameter $M_1$, we now fix the $\tau\,$-Yukawa coupling 
to $h_\tau = 0.007$, close to its physical value, and vary 
$M_1\rightarrow \alpha M_1$ instead. To keep the effect of the
washout term and the source term constant, we also scale 
$Y_{11} \rightarrow \sqrt{\alpha} \,Y_{11}$ and $Y_{12} \rightarrow \sqrt{\alpha}\, Y_{12}$ as well as $M_2 \rightarrow \alpha M_2$. This scaling behaviour can be seen when recasting
Eq.~(\ref{kin:eq}) into the form
\begin{align}
z H\,
\frac{\partial q_{\ell ab}}{\partial z}
=
\frac1a
\left\{
%-{\rm i}\Delta\omega^{\rm eff}_{\ell ab} q_{\ell ab}
\sum\limits_{c}\left[
q_{\ell ac}\Xi_{cb}
-\Xi_{ac}q_{\ell cb}
-W_{ac} q_{\ell cb}
-q_{ac}W_{cb}
\right]
+ 2 S_{ab}
-\Gamma_{\ell ab}^{\rm fl}
\right\}
\,.
\end{align}
The terms on the right hand side now correspond to the physical instead of
conformal interaction rates per unit volume.
Since at fixed $z$, $T/a(\eta)\to \alpha T/a(\eta)$, $H\to \alpha^2 H$ and
$q_\ell\to \alpha^3 q_\ell$, both sides scale as $\alpha^5$, except for the term
$1/a\times\Gamma^{\rm fl}_\ell$, which scales as $\alpha^4$. Therefore,
all the scale-dependence is isolated within the flavour-dependent damping rate.

\begin{figure}
\begin{center}
%\begin{tabular}{c}
\epsfig{file=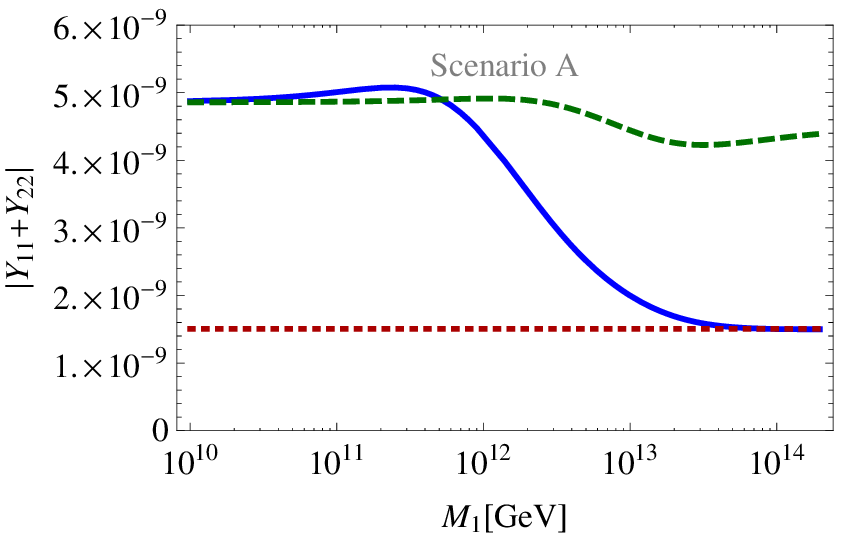,width=12.cm}
%\hskip.5cm
\\
% %(A)
% \end{tabular}
% \begin{tabular}{c}
\epsfig{file=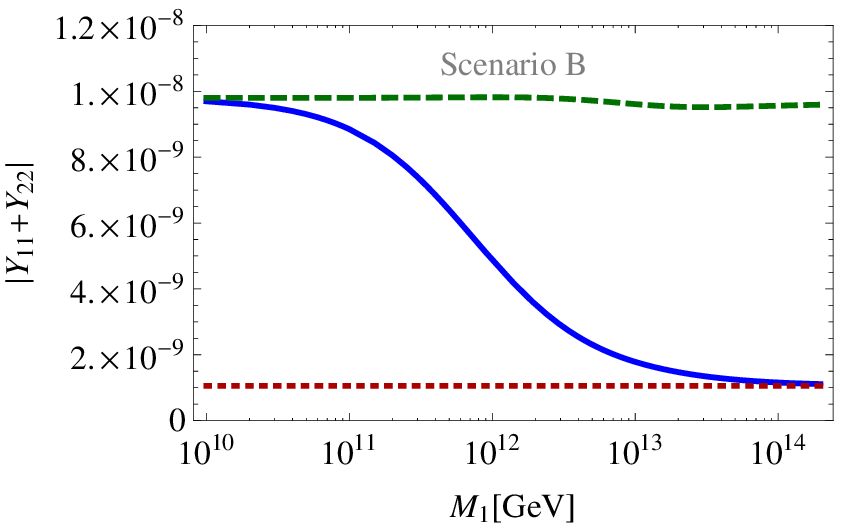,width=12.cm}
%\\
%(B)
%\end{tabular}
\end{center}
\caption{
\label{fig:total}
Shown is the total lepton asymmetry ${\rm tr}[Y_\ell] = Y_{\ell 11} +
Y_{\ell 22}$ as a function of the righthanded neutrino mass $M_1$, for
parameters corresponding to Scenarios~(A) and ~(B). The result of the
full kinetic equations including flavour effects (solid blue line) is
compared to the results of the unflavoured approximation $h_\tau=0$ 
(red, dotted) and to the results from a fully flavoured approximation that neglects off-diagonal flavour excitations (green, dashed). 
}
\end{figure}

We solve the kinetic equations for $ 10^{10}\,{\rm GeV} < M_1 < 2\times 10^{14}\,{\rm GeV}$. Parametrically this brings us from a regime where flavour effects are maximal to the unflavoured regime~\cite{Blanchet:2006ch}. For comparison we also calculate the lepton asymmetry over this parameter range using first the unflavoured approximation ($h_\tau=0$)
and then using the fully flavoured approximation, where the off-diagonal number densities are set to zero throughout the calculation. 

The results are shown in Figure~\ref{fig:total}. We find that both the fully flavoured approximation and the unflavoured approach lead to accurate predictions of the total lepton asymmetry within their expected ranges of validity. The intermediate regime where the full kinetic equation needs to be solved ranges from around $5 \times 10^{11}\,{\rm GeV} - 10^{13}\,{\rm GeV}$ for Scenario~(A) and even further for Scenario~(B) where the unflavoured behaviour is only recovered for $M_1 \stackrel{>}{{}_\sim}10^{14}\,{\rm GeV}$.
This is because the condition for the
unflavoured description to be valid,
$\Gamma^{a}_{\rm ID}
\stackrel{>}{{}_\sim} h_{\tau}^2(\Gamma^{\rm an}+\Gamma^{\rm sc})$
for $a=1,2$ is only fulfilled for larger values of $M_1$ within Scenario~(B).
Besides, within the flavoured approximation of
Scenario~(B), the flavour $a=2$
is only weakly washed out. Therefore, quantitatively relevant contributions to the lepton asymmetry arise at earlier times,
where $\Gamma^a_{\rm ID}/[h_{\tau}^2(\Gamma^{\rm an}+\Gamma^{\rm sc})]$
is enhanced compared to this ratio close to freeze-out. As a consequence,
the fully flavoured description of Scenario~(B) requires smaller values for $M_1$
when compared to Scenario~(A).
Note that the absolute limits for the validity of the unflavoured or fully flavoured description may vary by up to order one factors due to the uncertainty in the overall prefactor of $\Gamma^{\rm fl}$. We identify
this also as a probable source of the numerical difference between the present work and
Ref.~\cite{Blanchet:2006ch}, where it was found
using $\Gamma^{\rm an}+\Gamma^{\rm sc}\approx 5\times 10^{-3} \,T/a(\eta)$
that the unflavoured description is valid already 
for $M_1 \geq 5\times 10^{11}\;{\rm GeV}$. 

\section{Conclusions}
\label{sec:conclude}

Using the CTP formalism, we have derived and solved kinetic equations that
describe flavoured leptogenesis. Our results allow for systematic calculations
of the lepton asymmetry within the intermediate regime, that is neither fully flavoured
nor unflavoured, and where off-diagonal correlations between the lepton densities are
of importance. The CTP framework proves particularly suitable for this problem,
since off-diagonal densities are straightforwardly implemented within the two-point
Green functions.

So far, kinetic equations that describe flavoured leptogenesis within the intermediate
regime have only been available as extrapolations from a toy model description of
number density matrices in the occupation number formalism~\cite{Abada:2006fw}.
The importance of a more systematic derivation of these equations
has been emphasised for example in Ref.~\cite{Blanchet:2006ch}.
Our main result, Eq.~(\ref{kin:eq}), that is
derived within the CTP framework with the Lagrangian~(\ref{TheLagrangian})
as the starting point, turns out to
resemble the corresponding equation of Ref.~\cite{Abada:2006fw} in many details of its
flavour structure, but it also exhibits qualitative differences.
The main improvements provided with the present work may be summarised as follows:
\begin{itemize}
\item
We derive the impact of the thermal dispersion relations on
the kinetic equations for the left-handed leptons. This confirms the
expectation that the effects of these dispersion relations share some qualitative features
with flavour oscillations induced by tree-level mass terms and allows for
quantitative predictions. Our results might also be useful for describing the dynamics
of neutrino flavours in interacting backgrounds.
\item
We find that fast pair creation and annihilation processes through gauge interactions
effectively overdamp flavour oscillations. This is a qualitatively distinct feature
from the results of Refs.~\cite{Abada:2006fw,De Simone:2006dd}.
\item
The washout, source and damping terms in the kinetic equations are derived from first
principles. They correspond to collision terms which we explicitly present in the
form of integrals. Some of these
integrals crucially depend on finite width effects, which make their evaluation difficult.
For the purpose of our numerical examples, we
make only estimates for those collision terms that strongly depend on finite-width
effects. Yet, these collision terms are well-defined, and their quantitatively accurate
evaluation may be the subject of future work, possibly using the methods of
Ref.~\cite{Arnold:2000dr}.
\end{itemize}

For a more accurate prediction of the lepton asymmetry, the present
analysis has to be
supplemented by a number of improvements, which vary in how detailed
they have been discussed in the literature yet and in how straightforwardly the present
work can be generalised to include them.
First, there are the so-called spectator processes~\cite{Buchmuller:2001sr,Nardi:2005hs},
which are transitions induced by 
Yukawa couplings and by strong and weak sphalerons and which transfer charges between the
$\ell_a$ and $\phi$ to other particles of the Standard Model. Note that
also the scatterings induced by $h$, that transfer charges to the charged right-handed 
leptons belong to this category. Depending on whether the additional interactions are
fully equilibrated, out-of equilibrium or in an intermediate regime, the kinetic 
equations have either to be supplemented by algebraic constraints for the various 
charges, or the network of equations has to be extended in a way that is similar to how we
account for the charged right-handed leptons. Conceptually more interesting
and challenging is the systematic inclusion of thermal effects. In the present work,
we have noted that the finite width and thermal mass effects allow for certain three-body 
processes that are kinematically forbidden in the vacuum. Again, the CTP formalism
bears the potential to account for thermal effects more systematically and may hence
serve to confirm or to extend earlier results on these effects~\cite{Giudice:2003jh}.
Therefore,
important improvements remain to be incorporated in order to calculate the lepton
asymmetry of the Universe to a good accuracy and
with a quantifyable account of the theoretical uncertainty. In order to achieve this 
goal,
the systematic computation of the flavour effects from first principles as presented in
this work may serve as a building block.

\subsubsection*{Acknowledgements}

\noindent
This work is supported by the Gottfried Wilhelm Leibniz programme 
of the Deutsche Forschungsgemeinschaft and by the Schweizer Nationalfonds.

\begin{appendix}
\numberwithin{equation}{section}
\section{Pole-Mass Equation and Finite Width Propagators}
\label{appendix:finitewidth}

In this Appendix, we show how close to equilibrium, the
constraint equation~(\ref{eq:constraint}) and the
equations for the retarded and advanced propagators~(\ref{polemass:gradexp})
can be solved consistently to first order in gradients.
The first order corrections account for modified dispersion relations and
for finite widths.

In thermal equilibrium, the collision term is vanishing
on the right-hand side of the constraint equation~(\ref{eq:constraint}).
Since the collision term is already first order in gradients, we may therefore
neglect it within the constraint equations when being close to equilibrium.
Furthermore,
the propagators and self-energies are approximately flavour-diagonal, such that
we can express the constraint equation~(\ref{eq:constraint}) in the
simple form
\begin{align}
\label{eq:constraint:eq:1storder}
\left\{
k\!\!\!/-{\Sigma\!\!\!\!/}_\ell^H,
{\rm i} S^{<,>}_\ell
\right\}
-\left\{
{\Sigma\!\!\!\!/}^{<,>}_\ell,
{\rm i} S^H_\ell
\right\}
=0\,.
\end{align}

The spectral function $S^{\cal A}_{\ell}$ is defined
through Eq.~(\ref{CTP:combinations}).
In order to solve the pole-mass equation~(\ref{polemass:gradexp}),
it is useful to introduce
\begin{align}
\tilde{\Sigma\!\!\!\!/}_\ell=
\left(
\begin{array}{cc}
0 & \Sigma_\ell\cdot\sigma\\
\Sigma_\ell\cdot\bar\sigma & 0
\end{array}
\right)\,,
\end{align}
where $\Sigma_\ell^\mu=\frac12 {\rm tr}\gamma^\mu {\Sigma\!\!\!\!/}_\ell$.
This facilitates the inversion of propagators within the
four component formalism in analogy with the doubling of degrees
of freedom within the Weyl-fermion propagators that is familiar
from the in-out framework~\cite{Denner:1992me}.
We obtain ({\it cf.}~\cite{Garbrecht:2008cb})
\begin{align}
\label{spectral:function}
S_\ell^{\cal A}
=P_{\rm L}\frac{
2 \left(k\!\!\!/-\tilde{\Sigma\!\!\!\!/}_\ell^H\right)
\Sigma_\ell^{\cal A}\cdot(k-\Sigma_\ell^H)
-\tilde{\Sigma\!\!\!\!/}_\ell^{\cal A}\left(k\!\!\!/-\tilde{\Sigma\!\!\!\!/}_\ell^H\right)^2
+{{{\tilde{\Sigma\!\!\!\!/}}_\ell^{{\cal A}\,3}}} 
}
{
\left[
\left(k\!\!\!/-\tilde{\Sigma\!\!\!\!/}_\ell^H\right)^2
-{\tilde{\Sigma\!\!\!\!/}_\ell^{{\cal A}\,2}}
\right]^2
+4\left[
\Sigma_\ell^{\cal A}\cdot(k-\Sigma_\ell^H)
\right]^2
}
P_{\rm R}
\,.
\end{align}
Note again that we have used simplifications due to the flavour-diagonal
structure in equilibrium.
Similarly, we find the Hermitian propagator
\begin{align}
\label{S:hermitian}
S_{\ell}^H=
P_{\rm L}
\frac{
\left(k\!\!\!/-\tilde{\Sigma\!\!\!\!/}_\ell^H\right)
\left[
\left(k\!\!\!/-\tilde{\Sigma\!\!\!\!/}_\ell^H\right)^2
-{\tilde{\Sigma\!\!\!\!/}_\ell^{{\cal A}\,2}}
\right]
+2\tilde{\Sigma\!\!\!\!/}_\ell^{\cal A}
\Sigma_\ell^{\cal A}\cdot(k-\Sigma_\ell^H)
}
{
\left[
\left(k\!\!\!/-\tilde{\Sigma\!\!\!\!/}_\ell^H\right)^2
-{\tilde{\Sigma\!\!\!\!/}_\ell^{{\cal A}\,2}}
\right]^2
+4\left[
\Sigma_\ell^{\cal A}\cdot(k-\Sigma_\ell^H)
\right]^2
}
P_{\rm R}
\,.
\end{align}
Now, if we use that in equilibrium
\begin{align}
\vartheta(k^0)f^{{\rm eq}+}_{\ell ab}(\mathbf k)
-\vartheta(-k^0)(\mathbbm{1}_{ab}- f^{{\rm eq}-}_{\ell ab}(\mathbf k))
=\delta_{ab}\frac{1}{{\rm e}^{\beta k^0}+1}
\end{align}
and the KMS relation to substitute
\begin{align}
\label{SigmaA:Eq}
\tilde{\Sigma\!\!\!\!/}_\ell^{\cal A}
=-\frac{\rm i}{2}({\rm e}^{\beta k^0}+1)\tilde{\Sigma\!\!\!\!/}^<_{\ell}
\,,
\end{align}
we find that Eq.~(\ref{Seq:less}) with $S_\ell^{\cal A}$ given by 
Eq.~(\ref{spectral:function}) indeed solves the constraint
equation~(\ref{eq:constraint:eq:1storder})
in equilibrium. In a similar way,
the same observation holds for Eq.~(\ref{Seq:greater}). A related discussion
can be found in Ref.~\cite{Prokopec:2003pj}. Note that
relation~(\ref{SigmaA:Eq}) establishes the connection between the finite
width and the collision term~(\ref{collision:term}), that controls how fast
a small perturbation
in the lepton density relaxes to its equilibrium value. The
zero-width approximation~(\ref{S^A:singular}) follows from the 
result~(\ref{spectral:function}) in the limit
$\Sigma_\ell^{\cal A}\cdot(k-\Sigma_\ell^H)\to 0$.

Close to the poles, where
$(k\!\!\!/-\tilde{\Sigma\!\!\!\!/}_{\ell aa}^H)^2=0$, the
Hermitian propagator $S_\ell^H$ is suppressed compared to
the spectral function
$S_\ell^{\cal A}$ by an additional factor $O(\Sigma^{\cal A}/k)$.
To first order in the gradient expansion and
in the narrow width limit, where
$\Sigma_\ell^{\cal A}\cdot(k-\Sigma_\ell^H)\ll {k^0}^2$,
we can therefore neglect the terms involving
$S_\ell^H$ in Eqs.~(\ref{eq:constraint:kin}).

Plugging the ansatz~(\ref{S:decomposition}) into the constraint 
equation~(\ref{eq:constraint:eq:1storder}) and neglecting the term
$\left\{
{\Sigma\!\!\!\!/}^{<,>}_\ell,
{\rm i} S^H_\ell
\right\}$
leads us to
\begin{subequations}
\label{constraint:equations}
\begin{align}
g_{h0}^{<,>} k^0 + h |\mathbf k| g_{h3}^{<,>}
-\frac 12
\left\{
\bar\varsigma^{\rm bl}+\bar \varsigma^{\rm fl},g_{h0}^{<,>}
\right\}
-\frac h2
\left\{
{\rm sign}(k^0)(\bar\varsigma^{\rm bl}+\bar \varsigma^{\rm fl})
-\varsigma^{\rm bl}-\varsigma^{\rm fl},
g_{h3}^{<,>}
\right\}&=0\,,
\\
g_{h3}^{<,>} k^0 + h |\mathbf k| g_{h0}^{<,>}
-\frac 12
\left\{
\bar\varsigma^{\rm bl}+\bar \varsigma^{\rm fl},g_{h3}^{<,>}
\right\}
-\frac h2
\left\{
{\rm sign}(k^0)(\bar\varsigma^{\rm bl}+\bar \varsigma^{\rm fl})
-\varsigma^{\rm bl}-\varsigma^{\rm fl},
g_{h0}^{<,>}
\right\}&=0
\,.
\end{align}
\end{subequations}
When neglecting the hole modes,
for leptons ($k^0>0$) the helicity $h=-1$ is negative, while for
anti-leptons ($k^0<0$) the helicity $h=1$ is positive.
In conjunction with the constraint~(\ref{Weyl:constraint}),
this implies within the flavour-diagonal basis the dispersion relations
\begin{align}
\label{eq:poles}
k^0=\pm\left[
|\mathbf k|+\varsigma^{\rm bl}+\frac12(\varsigma^{\rm fl}_{aa}+\varsigma^{\rm fl}_{bb})
\right]
\end{align}
for $g_{h ab}^{<,>}$.
Note that in the present case, $\varsigma^{\rm bl}\sim g_2^2$,
while  $\varsigma_{aa}^{\rm fl}\sim [h^\dagger h]_{aa}$.
Since $g^2_2 \gg [h^\dagger h]_{aa}$, the expressions for
the dispersion relations~(\ref{eq:poles}), that are accurate to order $g_2^2$,
are not reliable to order $[h^\dagger h]_{aa}$ in case $g_2^4 \stackrel{>}{{}_\sim} [h^\dagger h]_{aa}$.
However, since the flavour-blind terms are universal, the differences between
the dispersion relations
for different $i,j$ are nonetheless accurate to order $[h^\dagger h]_{aa}$ .

\section{Thermal Lepton Dispersion Relation Induced by the Right
  Handed Neutrino}
\label{appendix:thmass}

In this Appendix, we calculate the thermal correction to the dispersion relation of the left handed 
lepton $\ell$ induced by its Yukawa coupling $Y$ to the heavy right handed
neutrino $N_1$ and the Higgs boson $\phi$. The leading order contribution comes from the one-loop wave-function 
correction where $N_1$ and $\phi$ are running in the loop. A similar calculation has been performed in 
Ref.~\cite{Weldon:1982bn} for a massless fermion in the loop and in Ref.~\cite{Petitgirard:1991mf} for the
 light massive case. In the latter calculation, instead of the Higgs field, there is a massless
 gauge boson in the loop, but the intermediate results before taking the limit of small fermion mass can be
 straightforwardly applied to the present situation.

\subsection{Decomposition of the Self Energy}

The structure of the self energy and the dressed propagator is restricted by Lorentz invariance, which allows 
us to parametrise both by two invariant functions $a$ and $b$. These can be determined by an explicit 
calculation of the self energy as performed in detail in Ref.~\cite{Petitgirard:1991mf}, and they can be used 
to find the poles of the propagator and thereby the dispersion relation. We define $a$ and $b$ by:
\begin{equation}\label{PlasmaSelfEnergy}
{\Sigma\!\!\!\!/}^{H}_\ell
= \frac{1}{2}\left[{\Sigma\!\!\!\!/}^{T}_{\ell} + \gamma^0({\Sigma\!\!\!\!/}^{T}_{\ell})^\dagger\gamma^0 \right] = P_{\rm R}\left[-a\slashed{k} -b\slashed{u}\right]P_{\rm L}
\,.
\end{equation}
Using Eqs.~(\ref{thermal:mass},\ref{varsigma:nocoup},\ref{varsigmafl:decomp}) and (\ref{PlasmaSelfEnergy}) in the rest frame of the 
thermal bath with $u = (1,0,0,0)$, we can identify:
\begin{subequations}
\label{relation:sigma:ab}
\begin{align}
\bar{\varsigma}^{{\rm fl}} &= -k^0 a -b\,,\\
\varsigma^{{\rm fl}} &= -(|k^0| - |\mathbf{k}|)a-\text{sign}(k^0)b\,.
\end{align}
\end{subequations}
The poles of the propagators~(\ref{spectral:function}) and~(\ref{S:hermitian}) are given by the equation
\begin{align}
\left(k\!\!\!/- \Sigma\!\!\!\!/_\ell^H\right)^2 = 0\,,
\end{align}   
which determines the flavoured dispersion relation of the lepton $\ell$. Alternatively, the dispersion relation
 is given by the solution of the constraint equations~(\ref{constraint:equations}) leading to Eq.~(\ref{eq:poles})
 to first order in gradients (in particular, when assuming $a \ll 1$, which is justified
by perturbatively small coupling constants and loop suppression factors).
Substituting the relations~(\ref{relation:sigma:ab}) we then obtain the following first order dispersion relation in the flavour-diagonal basis:
\begin{align}
\label{dispersion:gen}
k^0 = \pm |\mathbf k| - \frac12(b_{{\rm D} aa}+b_{{\rm D} bb})
\,,
\end{align}
where $b_{\rm D} \equiv U^\dagger b U$. In accordance with Eqs.~(\ref{thermal:mass},\ref{varsigma:nocoup},\ref{varsigmafl:decomp}), $b$ can be decomposed into contributions from flavour blind gauge interaction and flavour sensitive ($h$- and $Y$-) Yukawa interactions:
\begin{align}
\label{b:decomposition}
b_{ab} = \delta_{ab} b^{\rm bl} + h^\dagger_{ac} h_{cb} b^{{\rm fl},h} + \sum\limits_i Y^*_{ia}Y_{ib} b^{{\rm fl},Y}_i\,,
\end{align}
with an analogous decomposition for $a$. We see that according to 
Eqs.~(\ref{dispersion:gen}), (\ref{b:decomposition}) different types of contributions to the full dispersion relation are simply additive. There is a caveat, however, if the contributions have a nontrivial hierarchy. For instance, in the heavy massive case with\footnote{In
order to simplify the notation, in
this Appendix, we denote by $T$ the physical temperature, in contrast to the comoving
temperature throughout the remainder of this paper.} $M_1 \gg T$, we
find that $b^{{\rm fl},Y}_1 \sim (T/M_1)^4$ while $a^{{\rm fl},Y}_1
\sim (T/M_1)^2$, {\it cf.} Eq.~(\ref{ab:heavy}) below. Then the term
mixing gauge- and $Y$-induced contributions, $b^{\rm bl}a^{{\rm
    fl},Y}_1 \sim T^4/(\mathbf{k}^2 M_1^2)$, 
would be of leading order in the $Y$-induced dispersion relation~(\ref{dispersion:gen}) instead of $b^{{\rm fl},Y}_1$, if $\mathbf{k}^2/M_1^2 < \frac{15 C(R)g^2}{128 \pi^2}$, where $C(R)$ is the quadratic Casimir of the fermion representation (see Ref.~\cite{Weldon:1982bn}) and $g$ is the gauge coupling. Using the values $C(R) = 3/4$ and $\alpha_{\rm W}= g^2/(4\pi) = 1/40$ and assuming $|\mathbf{k}| \sim T$, this condition reduces to $M_1 / T \gtrsim 10$.
The momentum region $|\mathbf{k}| \sim T$ is of particular relevance for leptogenesis,
since this is where most of the leptons $\ell$ are present.
For $z=M_1 / T \gtrsim 10$, the process of leptogenesis has
typically already completed. Thus we will not consider the case when
these mixed contribution dominate in this Appendix. Instead, we use
the additive dispersion relation~(\ref{dispersion:gen}) and consider
only the $Y$-induced contributions in what follows. We therefore 
subsequently also drop the superscript ``${\rm fl},Y$''.

The one-loop contribution to the thermal lepton self-energy induced by
the couplings $Y$ is given by 
\begin{equation}
 {\Sigma\!\!\!\!/}^{T,Y}_{\ell ab}(k)
=-{\rm i} Y^*_{1a}Y_{1b}
\int \frac{d^4p}{(2\pi)^4} {\rm i}\Delta^T_\phi(p)P_{\rm R}{\rm i}S^T_{N1}(p+k)\,,
\end{equation}
where ${\rm i}\Delta_\phi^T(p)$ and ${\rm i}S^T_{N1}(p+k)$ are the time-ordered thermal propagators for the 
Higgs field and the heavy right handed neutrino, respectively ({\it cf.} Ref.~\cite{Beneke:2010wd}). We 
have restricted the sum over the right-handed neutrino flavours to
$i=1$, since in the hierarchical case only 
the lightest right handed neutrino leads to a relevant contribution during leptogenesis.\footnote{More 
precisely, this restriction is justified below where we find that in the heavy massive case the leading order 
correction to the dispersion relation is proportional to $(T/M_i)^4$.} In Ref.~\cite{Petitgirard:1991mf}, only 
the light massive case is discussed where $T \gg M_1$. For leptogenesis however, we also need a calculation that 
is valid at late times, when $T \ll M_1$ as well as in the intermediate regime which can be treated only 
numerically. Since our calculation is very close to the one in Ref.~\cite{Petitgirard:1991mf} we do not repeat 
the first steps,\footnote{The difference in the calculations affects only the prefactors but not the structure 
of the integrations.} and continue evaluating the integrals given in Ref.~\cite{Petitgirard:1991mf} for the 
case of a heavy massive fermion in the loop.

\subsection{Heavy Massive Case}

In the heavy massive case $T \ll M_1$, we perform an expansion of the logarithmic functions 
$L^{B,F}_{\pm}(|\mathbf{p}|)$ defined  in Ref.~\cite{Petitgirard:1991mf} and of the neutrino number density 
$n_F(E)=1/({\rm e}^{E/T}+1)$ in powers of $T/M_1$. The energy of the heavy neutrino, $E(\mathbf{p})= \sqrt{\mathbf{p}^2+M_1^2}$, is of order $M_1$, and thus the distribution function $n_F(E)$ is exponentially 
suppressed for $T \ll M_1$. For this reason the fermionic contributions involving $L^{F}_{\pm}(|\mathbf{p}|)$ 
need not be considered. Within the functions $L^B_{\pm}$, we count $p^0$, $|\mathbf{p}|$ and $k^0$, $|\mathbf{k}|$ as order $T$, such that the expansion up to order $(T/M_1)^2$ yields\footnote{ $L^B_+(|\mathbf{p}|)$ 
has to be expanded one order further than $L^B_-(|\mathbf{p}|)$ to extract the correct result for 
$\text{tr}(\slashed{k}{\Sigma\!\!\!\!/}^{H,Y}_{\ell ab})$ up to a given order.}
\begin{align}\nonumber
L^B_+(|\mathbf{p}|) &= -\frac{8|\mathbf{p}|\,|\mathbf{k}|}{M_1^2}\left(1 + \frac{k^2}{M_1^2}\right)\,,\\
L^B_-(|\mathbf{p}|) &= 0\,.
\label{Logs}
\end{align}
Inserting Eqs.~(\ref{Logs}) into the expressions for the self-energy (in accordance with
Ref.~\cite{Petitgirard:1991mf}), the remaining integrations reduce to basic integrals and we obtain:
\begin{subequations}
\begin{eqnarray}
 \text{tr}(\slashed{k}{\Sigma\!\!\!\!/}^{H,Y}_{\ell ab}) &=&- Y^*_{1a}Y_{1b} \frac{k^2 T^2}{6M_1^2}\,,\\
 \text{tr}(\slashed{u}{\Sigma\!\!\!\!/}^{H,Y}_{\ell ab}) &=&- Y^*_{1a}Y_{1b} \frac{k^0 T^2}{6M_1^2}\,.
\end{eqnarray}
\end{subequations}
It is easy to relate these expressions to the $Y$-induced contributions to the Lorentz invariant functions $a$ and $b$ decomposed as Eq.~(\ref{b:decomposition}) to get
\begin{subequations}
\label{thermal:ab}
\begin{eqnarray}
a_{1} &=& \frac{T^2}{12 M_1^2}\,,\\ 
b_{1} &=& 0\,.
\end{eqnarray}
\end{subequations}
Up to order $(T/M_1)^2$ the dispersion relation (\ref{dispersion:gen}) then simplifies to the massless 
dispersion, $k^0= |\mathbf{k}|$. The calculation up to order $(T/M_1)^4$ requires a more tedious expansion. 
For the functions $a$ and $b$ we find
\begin{subequations}
\label{ab:heavy}
\begin{align}
a_{1} &= \frac{T^2}{12 M_1^2} \left( 1 + \frac{15 k^2 - 4 \pi^2 T^2}{15 M_1^2}\right)\,,\\ 
b_{1} &= \frac{4 \pi^2 k^0 T^4}{45 M_1^4}
\,.
\end{align}
\end{subequations}
At this order, the non-zero $b$ implies the following dispersion relation for
$g_{h ab}$ in the flavour-diagonal basis: 
\begin{align}
\label{disp:largeM1}
k^0 = {\pm} |\mathbf{k}|\left(1  - \tilde{Y}^2_{ab} \frac{4 \pi^2 T^4}{45 M_1^4} \right)\,,
\end{align}
where $\tilde{Y}^2_{ab} \equiv \frac12 \left(U^\dagger_{ac}Y^*_{1c}Y_{1d} U_{da} + U^\dagger_{bc}Y^*_{1c}Y_{1d} U_{db} \right)$, and $+$ applies to negative, $-$ to
positive helicity $h$.
Furthermore, we note that to the leading order $(T/M_1)^2$, Eqs.~(\ref{relation:sigma:ab}) and~
(\ref{thermal:ab}) imply that 
\begin{equation}
\varsigma^{{\rm fl},Y}_1 = - \frac{4 \pi^2 |k^0| T^4}{45 M_1^4}
\qquad \mbox{and} \qquad 
\bar{\varsigma}^{{\rm fl},Y}_1= -\frac{k^0 T^2}{12 M_1^2}\,. 
\end{equation}
Note also that $a$ is even in $k^0$ and $b$ is 
odd, such that $\bar{\varsigma}_1^{{\rm fl},Y}$ and $\varsigma_1^{{\rm fl},Y}$ have the correct symmetry 
properties~(\ref{symm:varsigma}). We emphasize that these results are valid for
\footnote{
Within this Appendix, we imply by $A\lesssim B$ that $A$ is either of order of $B$
or much smaller.
}
$|\mathbf{k}| \lesssim T \ll M_1$.

In the limit of $|\mathbf{k}| \gg M_1 \gg T$, we find another analytical expansion with 
\begin{align}
  \text{tr}(\slashed{k}{\Sigma\!\!\!\!/}^{H,Y}_{\ell ab}) =- Y^*_{1a}Y_{1b} \frac{T^2}{12}\,,
\end{align}
while the contribution from $\text{tr}(\slashed{u}{\Sigma\!\!\!\!/}^{H,Y}_{\ell ab})$ to the dispersion relation is vanishing in the leading order. The dispersion relation is then given by
\begin{align}
\label{disp:largeM1:2} 
k^0 = \pm|\mathbf{k}|\left(1 +  \tilde{Y}_{ab}^2\frac{T^2}{24 \mathbf{k}^2} \right)\,, 
\end{align}
from which an effective thermal mass for leptons $\ell$ with large momentum may be extracted:
\begin{align}
\label{mth:heavy}
m^Y_{\rm heavy} = \frac{|\tilde{Y}_{ab}|T}{2\sqrt{3}}\,.
\end{align}

\subsection{Light Massive Case}

We consider next the light massive case $M_1\ll T$. Assuming $|\mathbf{k}| \gtrsim T$ and further that the thermal corrections to the dispersion relation are small, $||k^0| - |\mathbf{k}|| \ll T$, we find that to the leading order, $b$ is given by
\begin{align}
\label{b:light}
b_{1} = \frac{T^2}{16\mathbf{k}^2}\left(\frac{k^2}{2|\mathbf{k}|}\text{ln}\left(\frac{k^0+|\mathbf{k}|}{k^0-|\mathbf{k}|}\right)-k^0\right)\,
\end{align}
Then, by using the assumption of small thermal corrections (in particular $a \ll 1$) we can use Eq.~(\ref{dispersion:gen}) to find the dispersion relation
\begin{align}
\label{disp:light}
k^0  = \pm |\mathbf{k}|\left(1 + \tilde{Y}_{ab}^2 \frac{T^2}{16 \mathbf{k}^2}\right)\,,
\end{align} 
where we have neglected a logarithmic correction proportional to $\left(\tilde{Y}_{ab}^2 \frac{T^2}{16 \mathbf{k}^2}\right)^2 {\rm ln}\left(\tilde{Y}_{ab}^2 \frac{T^2}{32 \mathbf{k}^2}\right)$. Now we can extract an effective thermal mass for $|\mathbf{k}| \gtrsim T$ by using the relation ${m^Y}^2 = {k^0}^2 - \mathbf{k}^2$ to find
\begin{align}
\label{mth:light}
m^Y_{\rm light} = \frac{|\tilde{Y}_{ab}|T}{2\sqrt{2}}\,.
\end{align}

For leptogenesis, where most of the leptons are within the momentum region $|\mathbf{k}| \sim T$, the results ~(\ref{disp:largeM1}) and~(\ref{disp:light})
imply that the matrix $U$ which diagonalises the lepton mass will be approximately 
constant for $T \gg M_1$ as all contributions to the dispersion relations
of $\ell$ are proportional to $T^2$. Then it is time dependent around $T \approx M_1$ and again 
approximately time independent when $T \ll M_1$, when the contributions induced by the couplings $Y$ are 
suppressed by $(T/M_1)^4$.

\subsection{Numerical solutions}

In this Section we solve the $Y$-induced dispersion relation numerically and compare it to the analytical
approximations (\ref{disp:largeM1}), (\ref{disp:largeM1:2}) and (\ref{disp:light}). For comparison, we also 
derive the dispersion relation and the thermal mass for the light massive case using hard thermal loop (HTL) 
approximation when $k^0, |\mathbf{k}| \ll T$. In this Section we consider only the single flavour case and set 
$\tilde{Y}_{ab}^2 \equiv |Y_1|^2 = 1$ in all plots. Note also that we do not take
account of relevant contributions from gauge couplings, such that the numerical
results presented
in this Section should be considered as checks of the analytic limits and as a study of
a toy system in the absence of gauge interactions.

\begin{figure}[ht!]
\begin{center}
\epsfig{file=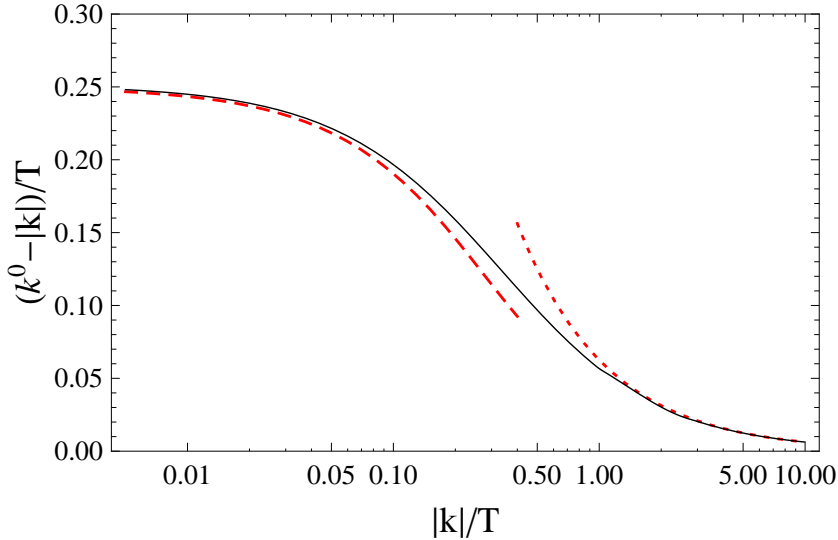,width=0.7\textwidth}
\end{center}
\caption{
\label{fig:dispersion:light}
Shown is the single flavour dispersion relation with $k^0 - |\mathbf{k}|$ as a function of $|\mathbf{k}|$ in the 
light massive case $M_1 \ll T$. The solid black line corresponds to the numerical solution, while the dashed red line to the approximate
dispersion relation~(\ref{disp:HTL}) and the
dotted red line to Eq.~(\ref{disp:light}).
}
\end{figure}

For the purpose of comparison to the numerical result, we give the analytical expression
for the dispersion relation for the light massive case $M_1 \ll T$ within the 
HTL approximation $k^0, |\mathbf{k}| \ll T$. Following Ref.~\cite{Petitgirard:1991mf} we find
\begin{subequations}
\begin{align}
 \text{Tr}(\slashed{k}{\Sigma\!\!\!\!/}^{H,Y}_{\ell ab}) &= Y^*_{1a}Y_{1b} \frac{T^2}{8}\,,\\
 \text{Tr}(\slashed{u}{\Sigma\!\!\!\!/}^{H,Y}_{\ell ab}) &= Y^*_{1a}Y_{1b} \frac{T^2}{16 |\mathbf{k}|}\text{ln}\left(\frac{k^0 +|\mathbf{k}|}{k^0 -|\mathbf{k}|}\right)\,,
\end{align}
\end{subequations}
which leads to
\begin{subequations}
\begin{align}
a_{1} &= \frac{T^2}{16\mathbf{k}^2}\left(1 - \frac{k^0}{2|\mathbf{k}|}\text{ln}\left(\frac{k^0+|\mathbf{k}|}{k^0-|\mathbf{k}|}\right)\right)\,, 
\label{HTL:a}
\\ 
b_{1} &= \frac{T^2}{16\mathbf{k}^2}\left(\frac{k^2}{2|\mathbf{k}|}\text{ln}\left(\frac{k^0+|\mathbf{k}|}{k^0-|\mathbf{k}|}\right)-k^0\right)\,.
\label{HTL:b}
\end{align}
\end{subequations}
From this expansion, we see that the suppression of $a$ due to small coupling
constants and loop suppression factors breaks down in the limit $|\mathbf k|\to 0$.
We notice that $b$ is given by the same expression (\ref{b:light}) as in the light massive case for $|\mathbf{k}| \gtrsim T$. 
This is not the case for $a$, however. To extract the thermal mass we expand $a$ and $b$ to the lowest order 
in $|\mathbf{k}|/|k^0|$ to get $a_{1} = -\frac{T^2}{48 {k^0}^2}$ and $b_{1} = -\frac{T^2}{24 k^0}$. Now we cannot 
use the dispersion relation (\ref{dispersion:gen}), since it has been derived assuming 
that $a \ll 1$, which is not the case when  $|\mathbf k|\to 0$. (Using Eq.~(\ref{mth:HTL}) below,
we find 
that $a = 1/3$ for $|\mathbf{k}|=0$). In the single flavour case, we find an identical dispersion relation to the one in 
Ref.~\cite{Petitgirard:1991mf}: $(k^0 \mp |\mathbf{k}|)(1+a) + b = 0$,
which now gives 
\begin{align}
\label{disp:HTL}
k^0 - |Y_1|^2 \frac{T^2}{16 k^0} = \pm |\mathbf{k}|\left(1 - |Y_1|^2 \frac{T^2}{48 {k^0}^2}\right)\,.
\end{align}
The thermal mass in the limit $|\mathbf{k}| \to 0$ is then given by
\begin{align}
\label{mth:HTL}
m^Y_{\rm HTL} = \frac{|Y_1|T}{4}\,.
\end{align}
This is a well-known result within the HTL approximation, see Ref.~\cite{Weldon:1982bn}.

\begin{figure}[p]
\begin{center}
\includegraphics[width=0.75\textwidth]{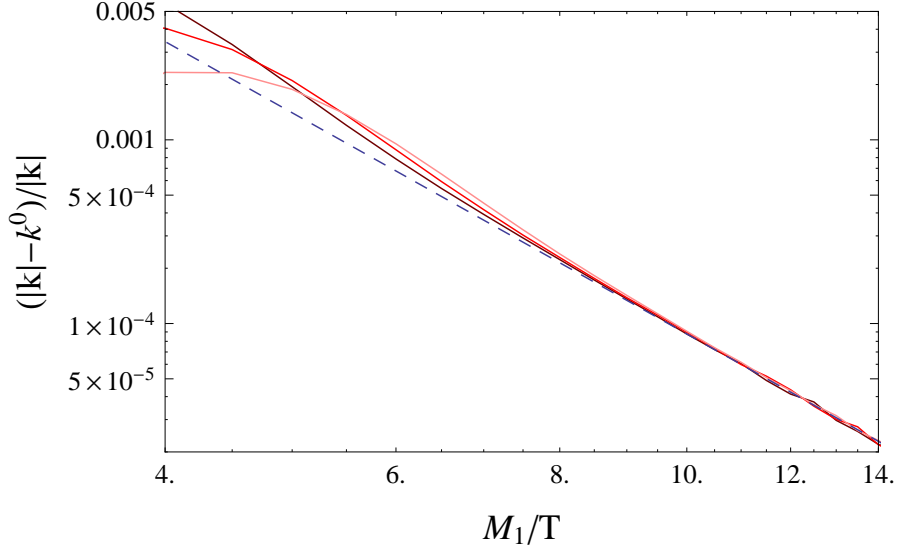}
\end{center}
\caption{
\label{fig:asymptotic:heavy}
Shown is the dispersion relation for the lepton $\ell$ with $|k^0 - |\mathbf{k}||/|\mathbf{k}|$ as a function of $M_1/T$ for $|\mathbf{k}|/T = 0.7, 1.0, 1.3$ from dark red (top at left) to light red (bottom at left). The dashed line correspond to the analytic approximation (\ref{disp:largeM1}).
}
\end{figure}
\begin{figure}[p]
\begin{center}
\vskip0.3cm
\includegraphics[width=0.7\textwidth]{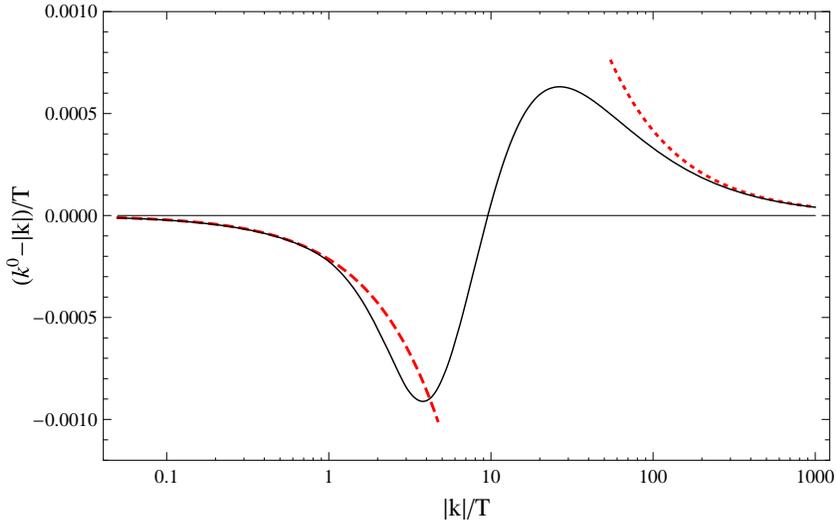}
\end{center}
\caption{
\label{fig:dispersion:heavy}
Shown is the dispersion relation with $k^0 - |\mathbf{k}|$ as a function of $|\mathbf{k}|$ in the heavy massive case with $M_1/T = 8$. The solid black line corresponds to the numerical solution, while the red dashed and dotted lines are the approximate dispersion relations~(\ref{disp:largeM1}) and~(\ref{disp:largeM1:2}). Also shown is the light cone $k^0 = |\mathbf{k}|$ (horizontal line).}
\end{figure}
\begin{figure}[ht!]
\begin{center}
\epsfig{file=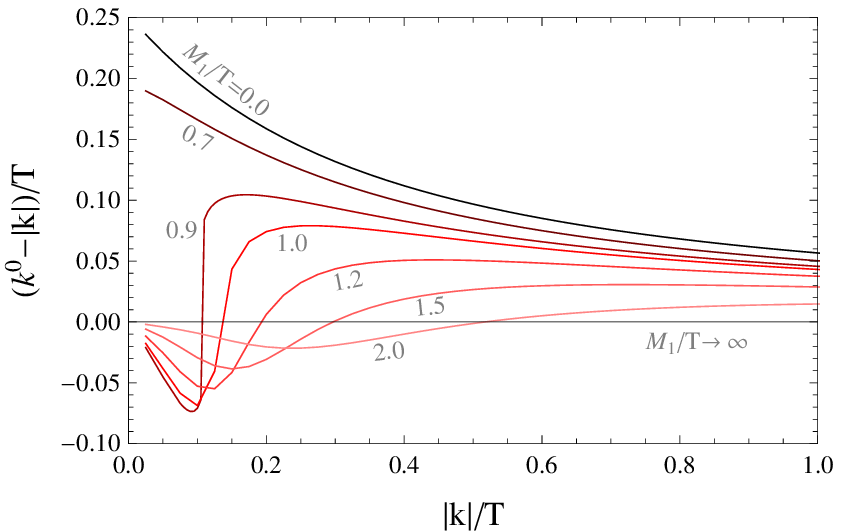,width=0.8\textwidth}
\end{center}
\caption{
\label{fig:dispersion}
Shown numerically obtained dispersion relations with $k^0-|\mathbf{k}|$ as a function of $|\mathbf{k}|$, for $M_1/T = 0.7, 0.9, 1.0, 1.2, 1.5, 2.0$ from dark red (top) to light red (bottom). Also shown is the case with $M_1/T = 0$ (black, top) and the light cone $k^0 = |\mathbf{k}|$ (black, middle). 
}
\end{figure}
%
%The interesting parametric region for leptogenesis is, however, $|\mathbf{k}| \sim T$, since most of the leptons a%re within this momentum region. 

In Figure \ref{fig:dispersion:light}, we compare the approximate dispersion relations~(\ref{disp:light}) and~(\ref{disp:HTL}) for the light massive case $M_1 \ll T$ to the exact numerical solution. We confirm that the HTL approximation agrees with the exact numerical solution only for small $|\mathbf{k}|$, while approximating the dispersion by the thermal mass~(\ref{mth:light}) is reasonably accurate in the region $|\mathbf{k}| \gtrsim T$.

We next turn to heavy massive case $M_1 \gg T$. In Figure \ref{fig:asymptotic:heavy} we compare the large 
$M_1/T$ limit of the approximate analytical dispersion relation (\ref{disp:largeM1}) to the numerical 
solutions to find a reasonably good agreement for $M_1/T \gtrsim 10$. In addition,
these plots clearly confirm the asymptotic $(M_1/T)^4$ behaviour.

Moreover, we notice that the approximate dispersion relations (\ref{disp:largeM1}) and (\ref{disp:largeM1:2}) 
give $|k^0| < |\mathbf{k}|$ for small $|\mathbf{k}|$ and $|k^0| > |\mathbf{k}|$ for large $|\mathbf{k}|$, so 
we conclude that in between their regions of validity the dispersion curve must cross the light cone $k^0=|\mathbf k|$, implying 
an intermediate region of superluminal group velocity $v_g = \frac{d |k^0|}{d |\mathbf{k}|} > 1$,
which might indicate dissipative effects in this momentum region. This interpolation is indeed 
confirmed by the exact numerical solution of the dispersion relation. In Figure~\ref{fig:dispersion:heavy} we 
show the approximate dispersion relations together with exact numerical solution for $M_1/T = 8$. 
We see that the analytic approximations are fairly accurate in the regions $|\mathbf{k}| \lesssim T$ and 
$|\mathbf{k}| \gg T$ and the crossing of the light cone takes place at the momentum $|\mathbf{k}| \sim 10 T$.

The region in between large and small $M_1/T$ can be treated only numerically. In Figure \ref{fig:dispersion} we 
show the numerical solution for the dispersion relation for several values of $M_1/T$. We observe that 
for $M_1/T \geq 0.9$ the dispersion curves cross the light cone at larger $|\mathbf{k}|$ for larger $M_1$. It would be of interest
to investigate how finite-width effects and
thermal masses for the neutrino and the Higgs fields affect
the present analysis.

\end{appendix}

\end{document}